\documentclass[%
 reprint,
superscriptaddress,
 amsmath,amssymb,
 aps,
 prx,
]{revtex4-1}

\usepackage{graphicx}
\graphicspath{ {figures/} }
\usepackage{dcolumn}
\usepackage{bm}
\usepackage{color,soul}
\usepackage{float}

\def\cO{\mathcal{O}}

\global\long\def\ket#1{\left|#1\right\rangle }%

\global\long\def\bra#1{\left\langle #1\right|}%

\global\long\def\ketbra#1#2{\ket{#1}\bra{#2}}%

\begin{document}

\title{Propagation of errors and quantitative quantum simulation with quantum advantage}

  \author{S. Flannigan}
 \thanks{S.F. and N.P contributed equally to this work.} \affiliation{Department of Physics and SUPA, University of Strathclyde, Glasgow G4 0NG, UK}

  \author{N.~Pearson}
      \thanks{S.F. and N.P contributed equally to this work.}
      \affiliation{Department of Physics and SUPA, University of Strathclyde, Glasgow G4 0NG, UK}
    \affiliation{Center for Quantum Devices, Niels Bohr Institute, University of Copenhagen, 2100 Copenhagen, Denmark}\affiliation{Theoretische Physik, ETH Z\"urich, 8093 Z\"urich, Switzerland}
 
  \author{G.~H.~Low}
  \affiliation{Microsoft Quantum, Redmond, Washington 98052, US}
  \author{A.~Buyskikh}
  \affiliation{Department of Physics and SUPA, University of Strathclyde, Glasgow G4 0NG, UK}
  \affiliation{Riverlane, Cambridge CB2 3BZ, UK}
  \author{I.~Bloch}
  \affiliation{Ludwig-Maximilians-Universit\"at, 80799 M\"unchen,  Germany}
  \affiliation{Max-Planck-Institut f\"ur Quantenoptik, 85748 Garching, Germany}
  \affiliation{Munich Center for Quantum Science and Technology (MCQST), 85748 M\"unchen, Germany} 
  \author{P.~Zoller}
  \affiliation{Center for Quantum Physics, University of Innsbruck, Innsbruck A-6020, Austria}
  \affiliation{Institute for Quantum Optics and Quantum Information of the Austrian Academy of Sciences, Innsbruck A-6020, Austria}
  \author{M.~Troyer} 
  \affiliation{Microsoft Quantum, Redmond, Washington 98052, US}
  
  \author{A.~J.~Daley}
  \affiliation{Department of Physics and SUPA, University of Strathclyde, Glasgow G4 0NG, UK}

\date{\today}
\begin{abstract}
    The rapid development in hardware for quantum computing and simulation has led to much interest in problems where these devices can exceed the capabilities of existing classical computers and known methods. Approaching this for problems that go beyond testing the performance of a quantum device is an important step, and quantum simulation of many-body quench dynamics is one of the most promising candidates for early practical quantum advantage. We analyse the requirements for quantitatively reliable quantum simulation beyond the capabilities of existing classical methods for analogue quantum simulators with neutral atoms in optical lattices and trapped ions.  Considering the primary sources of error in analogue devices and how they propagate after a quench in studies of the Hubbard or long-range transverse field Ising model, we identify the level of error expected in quantities we extract from experiments. We conclude for models that are directly implementable that regimes of practical quantum advantage are attained in current experiments with analogue simulators. We also identify the hardware requirements to reach the same level of accuracy with future fault-tolerant digital quantum simulation. Verification techniques are already available to test the assumptions we make here, and demonstrating these in  experiments will be an important next step. 
\end{abstract}

\maketitle

State-of-the art analogue quantum simulators \cite{Cirac2012,Georgescu2014} are now at the stage where they reach between 50 and several thousand particles in optical lattices \cite{Bloch2012,RevModPhys.80.885,Choi2016,Mazurenko2017}, trapped ions \cite{Blatt2012,Zhang2017,Brydges2019}, or neutral atoms in tweezer arrays \cite{Browaeys2020,Bernien2017}, and explore both time-dependent dynamics and equilibrium properties of a range of lattice models (see Fig.~1a). At the same time, rapid development in hardware for quantum computing opens opportunities for digital quantum simulation (Fig.~1b) \cite{Lloyd1996,Childs9456,Muller2012}. An important question at this stage is to understand in which regimes the output of quantum simulators is quantitatively reliable. In this work, we analyse the error accumulation for current analogue quantum simulators, understanding the level of error we would expect from first-principles microscopic models, and also compare the gate count required for fault-tolerant digital quantum simulators to reach the same accuracy. These questions are particularly important when we operate beyond the capabilities of existing classical numerical techniques, and beyond exploring robust qualitative features. This also provides a potential definition for a \emph{practical quantum advantage}, when quantum simulators can -- with quantitative reliability -- solve a problem that is intractable to existing classical methods, and relevant (at least in science) beyond testing the hardware itself. 
  
  \begin{figure}[t]
    \includegraphics[width=8.5cm]{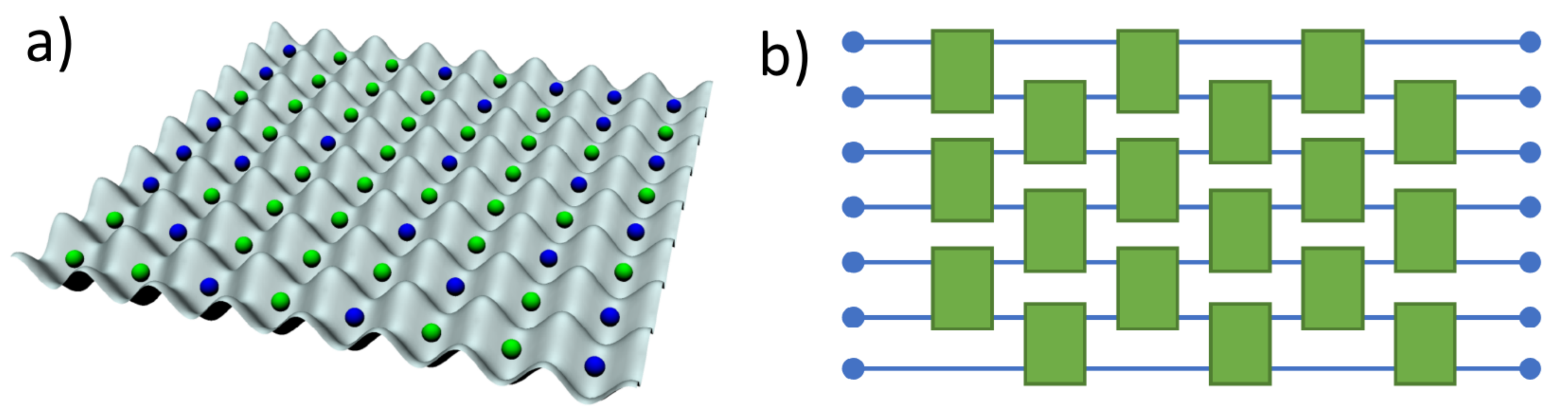}
    \centering
    \caption{Quantum simulation of time dependent dynamics can be performed either by (a) directly implementing the model to be simulated as the microscopic desciption (under well controlled approximations) on an analogue quantum simulator, or (b) computing the dynamics using a universal quantum computer as a digital quantum simulator.}
    \label{Fig1}
  \end{figure}

 For quantitative accuracy, we need to understand how the errors in measured results (even in an ideal case where we can reduce statistical measurement errors arbitrarily) depend on microscopic error sources relevant in experiments. These can include calibration errors, unwanted Hamiltonian terms, or effects arising from noise and decoherence. Though an analogue quantum simulator does not allow for arbitrary accuracy in general, for local models we expect that the propagation of local errors can be bounded and relatively small on relevant timescales \cite{poggi2020}, essentially because this is limited by Lieb-Robinson bounds \cite{Poulin2015,Kliesch2014,childs2019trotter}. The relative errors are then further suppressed if we aim to study intensive quantities (e.g., spatially averaged correlations in a translationally invariant case) on large systems. Quantum simulators based on neutral atoms or trapped ions then open a particular opportunity to analyse error propagation quantitatively, because these systems are typically well isolated from their environments, and we can derive microscopic models for coherent dynamics, dissipation, decoherence, and noise \cite{McKay:2011aa} under well-controlled approximations from first principles \cite{Daley:2014aa}.  Crucially, we note that the assumptions we make (both for the Hamiltonian and with noise and dissipation) can be tested by verification methods in experiments \cite{Carrasco2021a} – for example, via Hamiltonian \cite{li2020hamiltonian, Bairey2019, Evans2019, Valenti2019, Wang2017} or Liouvillian \cite{Pastori2022} Learning techniques.
 
 Below, we analyse errors occuring in current experimental platforms, evaluating the sources of error in the simulation of dynamics after a quench in the Hubbard \cite{leblanc2015,essler2005one} and transverse field Ising \cite{sachdev2011quantum} models. These are natural models to study because they are textbook examples from condensed matter physics, and their dynamics are generally difficult to compute, especially for inhomogeneous models, and especially in higher dimensions. We initially explore this for 1D systems in classically computable regimes. Because the limits on error propagation also apply in higher dimensions, which we explore by extending our calculations to ladder systems, 2D systems provide a further advantage for quantum simulators over classical methods. For existing analogue quantum simulators that implement these models, where we conclude that quantitatively reliable results are well within the capabilities of current experiments, in parameter regimes that are intractable to current classical methods.
 
We then make a comparison with the resources required for future digital quantum simulation \cite{Lloyd1996,Childs9456}. Digital quantum simulators will have two general advantages, firstly the potential to simulate a wider class of models than can be engineered directly an an analogue quantum simulator, and secondly, the potential (at least in principle) for arbitrary accuracy. However, the hardware requirements are often very significant, and beyond the current roadmaps of Noisy Intermediate-Scale Quantum (NISQ) devices. We thus analyse the number of qubits and the depth of the quantum circuit needed to perform the calculation with the same accuracy on fault-tolerant logical qubits as we obtain from our analysis of analogue quantum simulators. We show that the hardware requirements to reach this same point for digital quantum simulation are substantially reduced compared with previous studies on the same models, but still require substantial development of the hardware. 

 The rest of this manuscript is organised as follows. In section \ref{sec:comparison} we give an overview of the assumptions we make, and discuss the sources of error we need to account for in analogue quantum simulation, and in fault-tolerant digital quantum simulation of dynamics. 
We then introduce our example models in section \ref{sec:model} and calculate the propagation of errors for implementations of these examples on current analogue simulators \ref{sec:errors}, where we determine the accuracy of implementation required to reach the quantum advantage regime. We then discuss the time limits for propagation set by noise and decoherence in section \ref{sec:timelimit}. In section \ref{sec:digital} we then estimate the number of gate operations required for a fault-tolerant digital quantum computer to reach the same accuracy.

This is followed by several appendices with further technical information on our calculations, and comparison with other sources of error. In appendix \ref{app:classical} we discuss the limits of current classical simulation of dynamics in these models. In appendix \ref{app:calibration}, we discuss more details of the influence of calibration errors and the assumptions and parameter choices we make. In appendix \ref{app:othererrors}, we discuss the errors that can arise from other sources, such as additional unwanted terms in the Hamiltonian or error in preparation of the initial state, and show that these are significantly smaller than the errors arising from calibration uncertainties or decoherence. We then discuss the details for the comparison to digital quantum simulation, especially in our choice of Trotter decomposition in appendix \ref{app:trotter}. Finally, in appendix \ref{app:gatecount}, we describe how we reached the gate count estimates for the digital quantum simulation.

\section{Assumptions and context for comparing classical simulation with analogue and digital quantum simulation}

\label{sec:comparison}

When analysing errors and reliability of quantum simulators, as well as making comparisons between classical, analogue, and digital quantum simulation, we need to be clear about the choices we make. In this section we summarise our approach.

Firstly, we focus on quench dynamics, and extracting few-body correlation functions. While there is a lot of recent work discussing potential quantum advantage in ground state calculations of strongly interacting models, the capabilities and potential of near-future classical calculations are in many cases unclear, and under constant development. In contrast, there are strong results constraining the efficiency of state-of-the-art tensor network methods for computing out-of-equilibrium quench dynamics \cite{schuch2008,SCHOLLWOCK201196}. In such a case, the entanglement (expressed, e.g., as a von Neumann entropy of the reduced density operator) grows linearly \cite{Calabrese2005}, so that generic quench dynamics will require exponential resources to obtain first-principles classical results. We will choose an example where we have dynamics that entangle of the order of 100 spins or particles, which is well beyond the capabilities of current classical methods. The information that needs to be extracted for applications generally involves local or few-body correlations, so we focus on the accuracy with which such observables can be extracted, rather than on state fidelities (although we also compare with these in Appendix A). We note that this is immediately extendable to other systems such as string correlation functions, which have been measured with quantum gas microscopes~\cite{arxiv_String_order}.

From analysing Lieb-Robinson bounds for propagation of correlations \cite{Poulin2015,Kliesch2014,childs2019trotter} and general considerations of random operations \cite{poggi2020} it can be understood that propagation of local errors will be bounded in quantum simulation of local models. However, to verify the reliability of a quantum simulation, we need to quantitatively analyse the influence of errors under realistic conditions. As noted above, the advantage of the current atomic physics platforms in this regard, including cold atoms in optical lattices \cite{Bloch2012,RevModPhys.80.885,Choi2016,Mazurenko2017}, trapped ions \cite{Blatt2012,Zhang2017,Brydges2019,Monroe2021}, and neutral atoms in tweezer arrays \cite{Browaeys2020,Bernien2017}, is that we can derive the microscopic many-body models from first principles under well-controlled approximations \cite{Jaksch2005,Muller2012}. This includes understanding the imperfections and sources of noise and decoherence \cite{Daley:2014aa, McKay:2011aa}. By the nature of an analogue device, errors can arise from several sources, specifically (i) global and local calibration errors in the Hamiltonian parameters, (ii) imperfect implementation of the Hamiltonian (e.g., neglected higher order perturbations~\cite{Jaksch2005}),  (iii) heating due to time dependent noise on the Hamiltonian parameters~\cite{PhysRevA.87.033606,PhysRevA.86.051605} and (iv) decoherence~\cite{Daley:2014aa}, as well as state preparation and measurement errors that include readout of observables (v) \cite{Chiu251} and (vi) in the preparation of initial product states ~\cite{Mazurenko2017} . For Hubbard model simulations with atoms in optical lattices, calibration of parameters dominates the effects of static errors (i)-(ii), and so we will pay particular attention to these below, while estimating the magnitude of other errors from static sources in appendix \ref{app:othererrors}. The timescale for analogue simulation is then ultimately limited by heating and decoherence, (iii)-(iv) in the experiments, which we also analyse below.

To analyse the hardware requirements for digital quantum simulation of these models, we first choose a Trotter decomposition that for the timescale simulated introduces an error comparable to the leading source of errors in our analysis of analogue systems \cite{Lloyd1996,Poulin2015,Kliesch2014,childs2019trotter}. We then minimise the gate count for this decomposition both for a NISQ and fault-tolerant system to make a comparison between the digital and analogue hardware requirements at the quantum advantage point. The total error in the digital simulation will be the combination of Trotter decomposition and gate errors and so by proceeding in this way we have prioritised the Trotter error in order to find a lower bound for the number of gates required. This trade-off could be executed with a different priority which would increase the final circuits and gate counts produced.

There is a natural inequality in implying a direct comparison between digital and analogue quantum simulators, as digital devices are universal, and not restricted to certain classes of Hamiltonians. Nonetheless, the problems we have chosen, and important classes of related models are accessible on analogue devices, making this comparison for models that can be implemented in an analogue manner useful. We provide a Q\# code in order to facilitate extensions and adjustments using different optimisation strategies or different gate sets, as these are likely to result in some variety of resource requirements. We will see below that the prospects for universal digital quantum simulation in the medium term are very promising, and that for those classes of problem where analogue simulators exist with sufficiently small calibration errors, the quantum advantage point is already accessible with current technology.

\section{Models and classical calculations}
\label{sec:model}

As test cases we consider two models for many-body quantum systems, firstly the Hubbard model, relevant for the physics of strongly interacting electrons. This can be realised with ultracold atoms in optical lattices,  where it can be derived from first principles under well-controlled approximations \cite{PhysRevLett.81.3108,Esslinger2010}. The Hamiltonian is given by
\begin{equation} \label{Hubb_Mod}
H_H = -J \sum_{n,\sigma} \left( \hat{c}^{\dagger}_{n,\sigma} \hat{c}_{n+1,\sigma} + h.c. \right) + U \sum_n \hat{n}_{n,\uparrow} \hat{n}_{n,\downarrow},
\end{equation}
for $\hbar\equiv 1$ and where $\hat{c}^{\dagger}_{n,\sigma}$ is the creation operator for a fermion with spin $\sigma=\{\uparrow, \downarrow\}$ on site $n$, and $\hat{n}_{n,\sigma}=\hat{c}^{\dagger}_{n,\sigma}\hat{c}_{n,\sigma}$. Here, $J$ is the tunnelling amplitude, and $U$ is the on-site interaction energy for two electrons of opposite spins. 

Secondly we consider a transverse field Ising model for interacting spins, of the form 
\begin{equation} \label{Ising_Mod}
H_S = \sum_{n<m} \tilde{J}^{(n,m)} \hat{S}^n_{z} \hat{S}^m_{z} + B\sum_n \hat{S}^n_{x},
\end{equation}
where $ \hat{S}^n_z$ is a spin-z operator at site $n$. Here, $B$ corresponds to an applied transverse field, and $\tilde{J}^{(n,m)}$ describes the strength of Ising coupling between different spins. This form of the interactions arises naturally in laser-driven chains of ions in a Paul trap ~\cite{PhysRevLett.92.207901,Grass:2014aa,Monroe2021} with $\tilde{J}^{(n,m)} \approx J_{0}/|n-m|^{\alpha}$. 
In the following we consider the case of long-range interactions with $\alpha=2$, but compare to a nearest neighbour Ising model with $\alpha\rightarrow \infty$ in appendix \ref{app:calibration}.

For each of these models we will now use classical simulation techniques to model the propagation of errors. As noted above, we will explore propagation of calibration errors for analogue devices, as well as heating and decoherence. We perform calculations of quench dynamics beginning in a product state, using a Time Evolving Block Decimation (TEBD) algorithm, \cite{Daley_2004,PhysRevLett.93.076401,SCHOLLWOCK201196}, always converged to numerical precision several orders of magnitude better than the error values we are computing. As noted above, due to the growth of entanglement in time for these systems, these calculation methods become exponentially more expensive after short timescales - but the timescales accessible here are sufficient for us to understand how propagation of errors affects observables in our chosen models.

\section{Comparison of calibration Errors and Trotter Errors}

\label{sec:errors}

To analyse the effects of calibration errors, we sample the model parameters ($U$ and $J$ for the Hubbard model, and $J_0$ and $B$ for the Ising models) from a normal distribution, with a standard deviation, $\Delta$, given as a percentage of the mean value. Repeatedly selecting values for the model parameters in this way and simulating the system many times we find the average values of some typical observables. We compare the values of these observables to those from the target model, which corresponds to parameters with the mean value of these distributions. Experimentally, these parameters are often set globally, by some external laser for example, and so the errors can be considered homogeneous. With this in mind we take model parameters that are uniform over the entire system for a given simulation, but we have compared this to the case of local calibration errors and found only small quantitative differences of less than $10\%$ of the error value.

We directly compare these errors with those from a Trotter decomposition of the time evolution operator into sequences of two-site gates which act on nearest neighbouring sites, $U_{n,n+1}$. This approach is relevant for digital simulation, which we will discuss in more detail below. These decompositions first consist of breaking the evolution up into discrete time steps where the choice of further decomposing the single time step evolution can be carried out in various ways to minimise the error. The first main case that we have considered are of the form of a left-right sweep, \cite{PhysRevA.60.1956}, where the two site operators are applied sequentially from left-to-right (and then right-to-left) throughout the whole chain. The second case, which we only apply to the Hubbard model is a Suzuki-Trotter decomposition of the odd/even bonds, \cite{SCHOLLWOCK201196}, where first all odd bonds are evolved and then all the even. See appendix~\ref{app:trotter} for more details.
We have compared the $2$nd and $4$th order versions of these methods in the time step, $\tau$. For the $4$th order sweep methods there are several decompositions which have different ratios of errors to complexity \cite{PhysRevA.60.1956},  we use the version with the smallest errors, 
\begin{equation}\label{trot_disp}
\begin{split}
e^{-i H \tau} \approx (1)^T&(1)(1)^T(-2)(1)^T(1)^T(1)^T(1)^T \\\
&(1)(1)^T (1)(1)(1)(1)(-2)^T(1)(1)^T(1),
\end{split}
\end{equation}
where $(s)$ corresponds to a single left-to-right sweep with a time step $s\tau/12$ and $(s)^T$ to a right-to-left sweep, such that the total cumulated time step of all sweeps in Eq.~\ref{trot_disp} is $\tau$, see appendix~\ref{app:trotter} for more details.
For the Hubbard model there is the additional choice of further decomposing the two-site operator into a contribution from the tunnelling, $J$, and the onsite interactions, $U$, but we find that this results in a slightly larger error as shown in appendix~\ref{app:trotter}. As this further decomposition is more likely to be compatible with available digital quantum hardware we present the errors found using this decomposition in the main text.

\begin{figure}[t]
  \includegraphics[width=8.5cm]{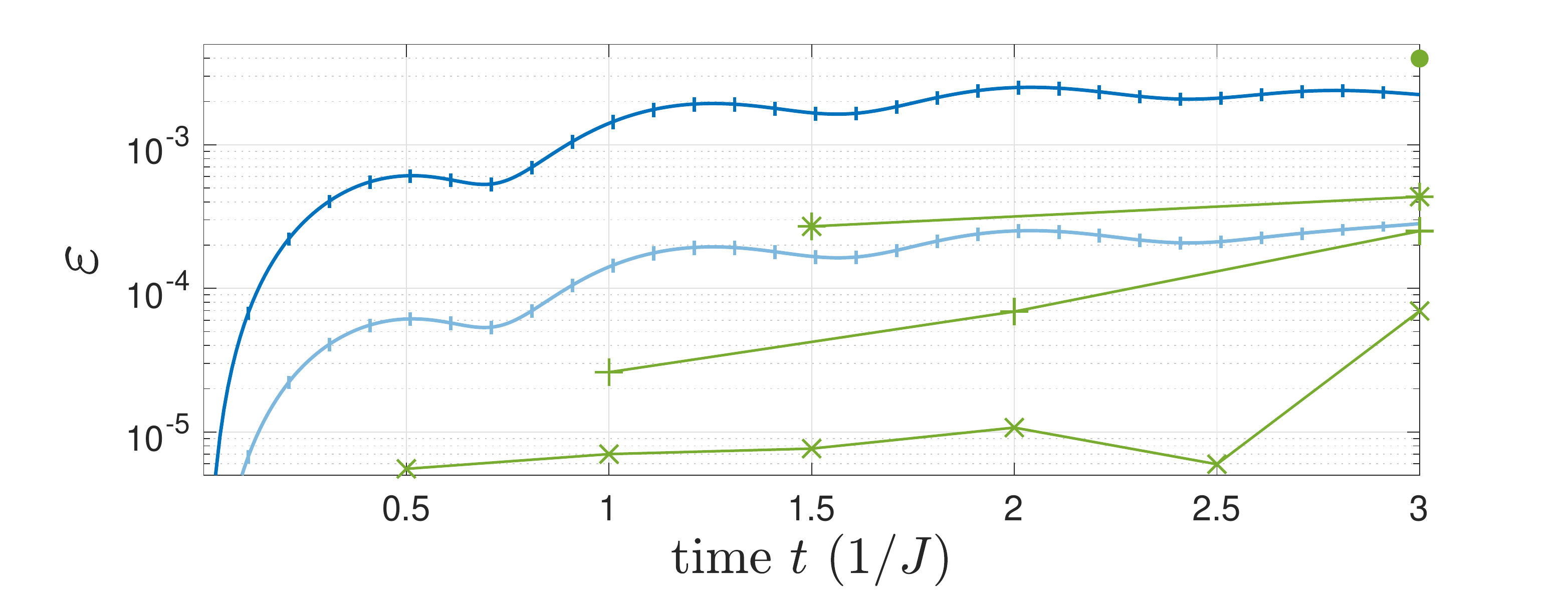}
  \centering
  \caption{Comparison of the RMS errors over all sites $n$ in the off-diagonal correlation function (Eq.~\ref{equ_error}) with $O_n = c^{\dagger}_{M/2,\uparrow} c_{n,\uparrow} $, for the Hubbard model, with system size $M=20$. For analogue simulation we compare a $1\%$ calibration error (dark blue) and a $0.1\%$ calibration error (light blue). For the digital we have plotted the 4th order decomposition (Eq.~\ref{trot_disp}), splitting  the interaction and kinetic energy terms, for the time steps $J\tau=3$ (circles), $J\tau=3/2$ (stars), $J\tau=1$ (plus sign) and $J\tau = 1/2$ (crosses).}
  \label{Fig2}
\end{figure}

\begin{figure}[t]
  \includegraphics[width=8.5cm]{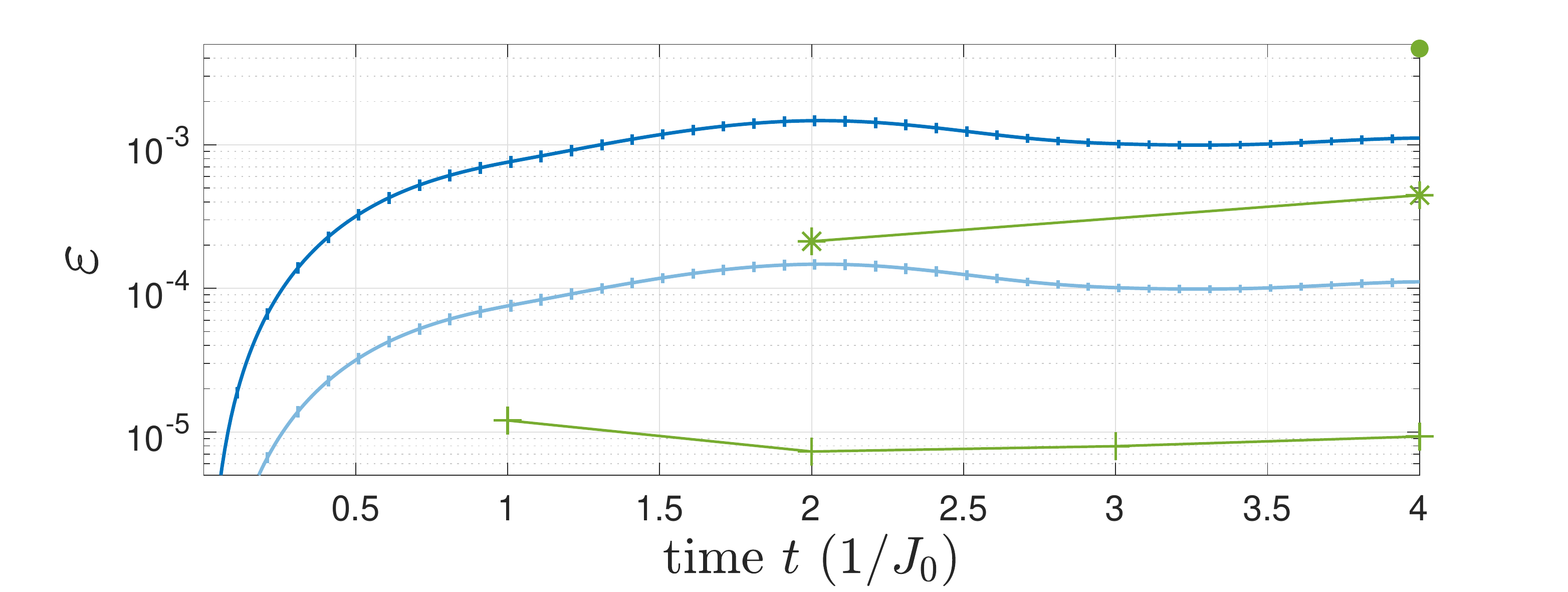}
  \centering
  \caption{Comparison of the RMS errors over all sites $n$ in the off-diagonal correlation functions (Eq.~\ref{equ_error}) with $O_n = S_+^{M/2} S_-^{n}$, for a Transverse field Ising model, with system size $M=20$. For analogue simulation we compare a $1\%$ calibration error (dark blue) and a $0.1\%$ calibration error (light blue). For the digital we have plotted the 4th order decomposition (Eq.~\ref{trot_disp}), for the time steps $J_{0}\tau=4$ (circle), $J_{0}\tau=2$ (stars), and $J_{0}\tau=1$ (plus sign).}
  \label{Fig3}
\end{figure}

In Fig.~\ref{Fig2} we plot the results of this analysis for the Hubbard model with $J=U=1$, beginning in the product state $|\Psi(0)\rangle = | \uparrow,\downarrow,\uparrow,\downarrow,\cdots\rangle$ and in Fig.~\ref{Fig3} for the long-range transverse Ising model, with $J_{0}=1=B=1$, beginning with all spins aligned in the $s_x$ direction,    $|\Psi(0)\rangle = | +,+,+,+\cdots\rangle$, for a system size, $M$, of $20$ lattice sites and spins respectively. We calculate the global errors in the observable, $O_{n}$ according to,
\begin{equation}\label{equ_error}
\varepsilon = \sqrt{\frac{1}{ M}\sum_n |\langle \psi_{\rm sim}(t) |  O_{n} | \psi_{\rm sim}(t) \rangle - \langle \psi_{\rm ex}(t) | O_{n} | \psi_{\rm ex}(t) \rangle |^2},
\end{equation}
where $| \psi_{\rm sim}(t) \rangle$ is the state as simulated on the experimental hardware with a given source of error and $| \psi_{\rm ex}(t) \rangle$ is the exact state. In Figs.~\ref{Fig2} and \ref{Fig3} we use the off-diagonal correlation functions,  $O_n = c^{\dagger}_{M/2,\uparrow} c_{n,\uparrow}$ and $O_n = S_+^{M/2} S_-^{n}$ respectively, as the observable and where the $4$th order left-right gate sweep is used for the digital simulation. We see that for the analogue simulation errors rapidly increase for short times but then remain approximately constant with some small oscillations and are at similar magnitudes in both the Hubbard model and Transverse Ising models. These oscillations are due to the calibration uncertainty introducing small differences in the frequency and amplitude of the oscillatory dynamical behaviour of the observables compared to the exact case. In addition, the average over different realisations of the parameters reduce the size of the oscillations observed in the errors. We note that that there is only a weak dependence on system size, as shown with additional calculations in appendix \ref{app:calibration}.

For the largest time classically simulable to this accuracy ($tJ=3$ for the Hubbard model and $tJ_{0}=4$ for the Ising model) we extrapolate the critical time step, $\tau^*$, defined as the time step for which the errors in the digital simulation will match or only slightly exceed the errors in the analogue simulation with a given level of calibration error. We find that the $4$th order decomposition for the Hubbard model has a critical time step of $J\tau^*\approx 2.7$ ($J\tau^*\approx 1$) in order for the errors in the observable correlation functions to match those arising from a calibration error of $\Delta=1\%$ ($0.1\%$), whereas for the Ising model $J_{0}\tau^*\approx 2.6$ ($J_{0}\tau^*\approx 1.5$) for a calibration error of $1\%$ ($0.1\%$). Note that these estimates for $\tau^*$ only weakly depend on system size and the particular observable, but depend more strongly on the type of Trotter decomposition used, up to at most a factor of 3 (see appendix \ref{app:trotter} for more data with different decompositions).


Finally, we performed similar simulations with a Hubbard model on a ladder system to confirm that there is no substantial quantitative change in the propagation of errors, and no change in the critical time step (see appendix~\ref{app:Extrap}). This implies that we can use the same time step for propagation of errors in a 2D system if we use the same period of time evolution as in 1D. The advantage of working with an extra dimension is that there is a much larger generation of entanglement in the larger system size.  Correlations spread at the same rate in the ladder system and a 1D chain, as is expected due to similar group velocities. Through this comparison we expect that the effect of calibration errors on our example correlation functions will be of the order of 1\% when extended to a 2D scenario, over timescales of $tJ=10$, which would be sufficient to entangle all parts of a $10\times10$ lattice system.

\section{Time limits for propagation in analogue simulators}

\label{sec:timelimit}

\begin{figure}[t]
  \includegraphics[width=8.5cm]{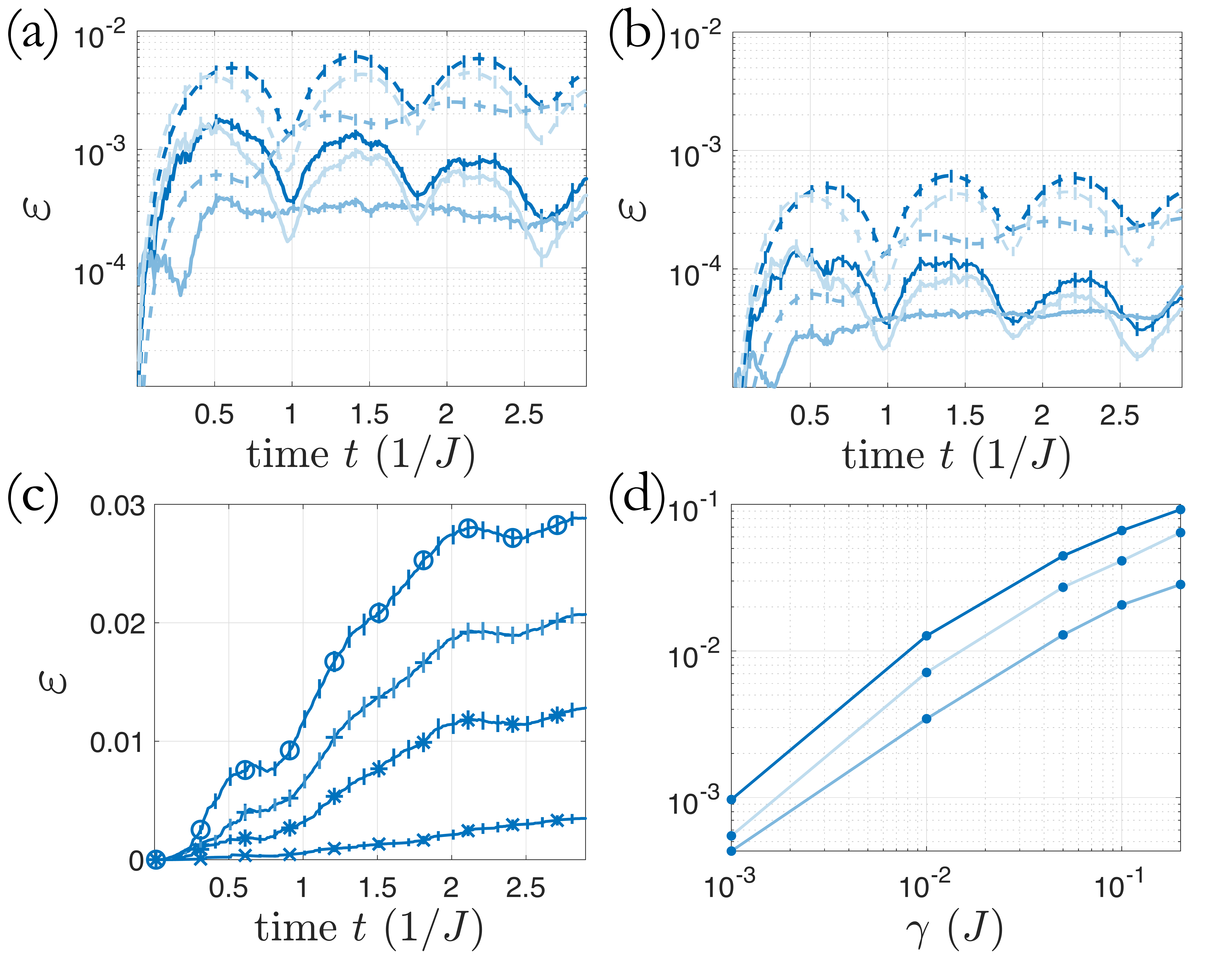}
  \centering
  \caption{(a-b) Comparison of the calibration errors (dashed) to the errors from laser fluctuations (solid) in the Hubbard model for different observables and magnitudes of error, $\Delta=\Delta_l = 1\%$ (a) and $\Delta=\Delta_l=0.1\%$ (b). We compare the errors, Eq.~\ref{equ_error}, in (from darkest to lightest) $O_n = c^{\dagger}_{n,\uparrow} c_{n,\uparrow} $, $O_n = c^{\dagger}_{M/2,\uparrow} c_{n,\uparrow} $ and $O_n = c^{\dagger}_{M/2,\uparrow} c_{M/2,\uparrow} c^{\dagger}_{n,\uparrow} c_{n,\uparrow} $. (c) Errors in $O_n = c^{\dagger}_{M/2,\uparrow} c_{n,\uparrow} $ due to decoherence with rate $\gamma/J=\{0.2,0.1,0.05,0.01\}$ (circles, plus signs, stars, crosses). (d) Errors due to decoherence at $tJ=3$ for different observables (from darkest to lightest) $O_n = c^{\dagger}_{n,\uparrow} c_{n,\uparrow} $, $O_n = c^{\dagger}_{M/2,\uparrow} c_{n,\uparrow} $ and $O_n = c^{\dagger}_{M/2,\uparrow} c_{M/2,\uparrow} c^{\dagger}_{n,\uparrow} c_{n,\uparrow} $. System size, $M=20$.}
  \label{Fig4}
  \end{figure}
  
The other error sources relevant to consider for analogue simulators are those of decoherence and heating which can create errors that grow in time, and that these will at some point contribute at the same level as calibration errors. These come from sources that are typically well characterised in experiments \cite{McKay:2011aa}
including spontaneous emission \cite{PhysRevA.82.063605,PhysRevA.82.013615}, and noise on the trapping lasers \cite{PhysRevA.86.051605,PhysRevA.87.033606}. First we incorporate the latter of these into simulations of the Hubbard model using time dependent parameters where the fluctuations in $U$ and $J$ are anti-correlated, since if the laser intensity increases, the potential barrier height increases which gives rise to a lower tunnelling rate but a larger onsite interaction strength. In Fig.~\ref{Fig4}(a-b) we compare errors caused by noise of trapping lasers to those produced from a calibration error of $\Delta=1\%$ \& $0.1\%$ of the parameters $U=J$. We directly compare these to the errors induced through laser fluctuations where we employ a white noise approximation~\cite{PhysRevA.86.051605,PhysRevA.87.033606} with a standard deviation given by $\Delta_l=1\%$ and $0.1\%$, where in this case, the percentage is given with respect to a recoil unit $E_R=\hbar^2k_l^2/2m$, with $k_l$ the wave-vector of the optical laser and $m$ the mass of the trapped atoms. We then calculate the parameters $U(t)$ and $J(t)$ from the resulting Wannier functions arising from the potential at each time step, choosing the mean intensity of the sinusoidal optical potential to be $V_0=10E_R$ (as this places us in a regime for typical experimental parameters where both $J$ and $U$ are significant), where the resulting tunnelling rate is calculated to be $J\approx 0.02 E_R$. We find that when we choose a distribution for the laser intensities with width $\Delta_l = 0.1\%$ of a recoil unit, then this gives rise to a tunnelling rate distribution with standard deviation $\sim 0.25 \%$ of a tunnelling rate, which means that in terms of overall uncertainty in energy, the chosen levels for the laser fluctuations are actually larger than the uncertainty in the calibration errors. The difference is that the fluctuations average to zero over time, so contribute to heating, whereas the calibration offset is seen to be constant over the period of one time evolution. In the figure the errors in each case have the same qualitative time dependence. However, those arising from the laser fluctuations are at a much lower overall magnitude compared to the effects from uncertainty in the calibration. 

We also analyse the effects of decoherence, which can arise through spontaneous emission by applying a quantum trajectory approach to the Lindblad master equation~\cite{Qnoise,Daley:2014aa} where we include independent jump operators at each site, $L_n = \sqrt{\gamma} \left( \hat{n}_{n,\uparrow} + \hat{n}_{n,\downarrow} \right)$. In Fig.~\ref{Fig4}(c-d) we plot the errors in observables where we have exaggerated the values of the decoherence rate, $\gamma$, in order to make the error scaling more explicit. For $\gamma=0.001J$, as is feasible in experiment, the level of error at the end of the evolution when $tJ = 3$ is comparable to errors in the same observables due to calibration uncertainty of $0.1\%$. State of the art analogue simulations quote coherence times on the order of $1-2~s$~\cite{PhysRevLett.124.203201,Deco_cite} which for $J\sim 1~{\rm kHz}$, correspond to values of $\gamma < 0.001 J$. This means that in realistic experiments the effects of decoherence will be smaller than those considered here for the same timescales. This analysis of decoherence and heating allows us to conclude that for the timescale we are interested in for the 2D system, these contributions can be neglected for typical experiments. 

We note that the combination of results in the previous two sections implies that - even at around 1\% calibration error, the dynamics we would observe at $tJ=10$ in a 10$\times$10 system of fermionic atoms in an optical lattice would result in an error of at most a few percent in the observables we have analysed. This is already beyond the capabilities of classical attempts to capture real time dynamics (see appendix \ref{app:classical}) and has notably already been realised in a series of experiments with quantum gas microscopes \cite{Boll2016,Cheuk2016,Parsons2016,Brown2017}).

\section{Digital circuit implementations and gate counts}
\label{sec:digital}

We now analyse the resources we would require to run this same calculation on a digital quantum computer, in order to reach the same error obtained for analogue quantum simulation. We note in in advance that for models that can be implemented on analogue quantum simulators, we the analogue implementations to have a natural advantage. However, future digital quantum simulators will allow arbitrary accuracy and a much wider range of models.

Having checked the cost of different Trotter decompositions, we choose that of Eq.~\ref{trot_disp}. We map the operations which execute a single time step to a set of gates which can be implemented on digital quantum computer. We assume a native gate set including single qubit rotations around the $z$ axis $R_z(\theta)=\textrm{exp}\{-i\frac{\theta}{2}\sigma_z\}$, Clifford gates and CNOT gates. Naturally the estimates and optimal decomposition are sensitive to the available gates but in most cases the two qubit gates and arbitrary rotations will have the most impact in limiting performance and are often common accross different hardware implementations.

For the Hubbard model of Eq.~\ref{Hubb_Mod} the standard way to do this mapping is to use the well known Jordan Wigner decomposition \cite{whitfield10}. We require $2M$ qubits to represent a system with $M$ sites and two species where the state of each qubit describes the occupation of one site by one species. Minimising the gate counts and depths for circuits of this kind has been a topic of extensive research \cite{hastings15}. Here we optimistically consider a digital quantum computer with all-to-all connectivity when calculating the gate counts and depths. While this may not be realistic for known qubit architectures, it will provide a lower bound for required number of qubits and runtime. To facilitate future estimates and allow for customization or improvements by other researchers we include a Q\# code \cite{Pearson_Quantum_Resource_Estimation_2022}, which we use to verify our gate count.

The on-site interaction term with energy $U$, along with any optional chemical potential terms, map easily to a circuit with $3M$ rotations (depth 2) and $2M$ CNOTs (depth 2). The main challenge is implementation of the tunneling terms with amplitude $J$, which consist of non-local Jordan Wigner strings of length $\cO(M)$ between the pairs of qubits involved in a tunneling event. This string consists of a chain of CNOT gates between the two qubits forming the pair. This necessarily involves all the intervening qubits which is expensive in CNOT gates and it is not immediately obvious how it might be possible to execute many of these in parallel. A key observation for optimising this circuit is that the Jordan Wigner string between qubits $k$ and $l$ effectively calculates the number of excitations between them. This calculation can be reused when finding the number of excitations between qubits $j$ and $m$ if $j<k<l<m$. Furthermore we note that executing the tunneling between $k$ and $l$ does \emph{not} change the parity on the string spanning $j$ to $m$ and so this operation can be commuted through the one executing the $j$ to $m$ tunneling event. This allows us to apply the rotation gates for both pairs in parallel. These considerations reduce the number of gates per time step sweep down from $\cO(M^5)$ to $\cO(M)$, with depths $\cO(\sqrt{M})$ at the expense of using $2\sqrt{M}$ additional ancilla qubits for storage of parity calculations \cite{wecker14, wecker15, hastings15}.

As a result the total number of rotation gates needed to implement one time step sweep is $11M - 8\sqrt{M}$ with depth $10$. The total number of CNOT gates is $2(15M - 20\sqrt{M} + 6)$ with depth $6(2\sqrt{M} + 1)$. Further details on the method for obtaining these counts can be found in appendix~\ref{app:gatecount} and the results for Hubbard model are shown below in Table~\ref{tab:hubbard}. 

\begin{table}[ht]
\centering
\begin{tabular}{ l | l | l }
  Gate & Gate Count & Depth\\ \hline 
  CNOT &  $1.7\times10^5$ & $8.4\times10^3$ \\
  $R_{Z}(\theta)$ & $6.8\times10^4$ & $6.7\times10^2$\\
\end{tabular}
\caption{Gate count and depth estimates for digital quantum simulation of the Hubbard model with $J\tau = 2.7$, $M=100$ and $tJ=10$.}\label{tab:hubbard}
\end{table}

Simulation of the Ising model of Eq.~\ref{Ising_Mod} is more easily mapped to operations on a digital quantum computer. Evolution under the transverse field maps to $M$ single qubit rotations with depth 1. Executing the Ising coupling term requires $ M(M//2)$ rotations and $2M(M//2)$ CNOT gates, with depths $M$ and $2M$ respectively. We ensure that the coupling terms between qubits are executed in parallel as much as possible to minimise the depth of the circuit. The resulting gate counts for the long-range Ising model are shown below in Table~\ref{tab:ising}.
\begin{table}[ht]
\centering
\begin{tabular}{ l | l | l }
  Gate & Gate Count & Depth\\ \hline 
  CNOT &  $6.9\times10^5$ & $1.4\times10^4$ \\
  $R_{Z}(\theta)$ & $3.5\times10^5$ & $7.0\times10^3$\\
\end{tabular}
\caption{Gate count and depth estimates for digital quantum simulation of the long-range Ising model with $J\tau=2.6$, $M=100$ and $tJ=10$.}\label{tab:ising}
\end{table}

If we instead set $\alpha\to\infty$ and estimate the resources of an Ising model with only nearest neighbour coupling then we need a significantly reduced number of coupling terms consisting of $2(M-1)$ rotations and $4(M-1)$ CNOT gates with depths only 4 and 8 respectively. This drastically reduces the overall resource estimates which can be seen in Table~\ref{tab:isingnn}. Here we have conservatively used the same critical time step as required for the long-range Ising model.

\begin{table}[ht]
\centering
\begin{tabular}{ l | l | l }
  Gate & Gate Count & Depth\\ \hline 
  CNOT &  $1.6\times10^3$ & $5.5\times10^2$ \\
  $R_{Z}(\theta)$ & $2.1\times10^4$ & $3.5\times10^2$\\
\end{tabular}
\caption{Gate count and depth estimates for digital quantum simulation of the nearest neighbour Ising model with $J\tau=2.6$, $M=100$, $tJ=10$.}\label{tab:isingnn}
\end{table}

For a computation with random error the final fidelity will be $(1-p)^N$ where $N$ is the number of gates and $p$ the error per gate. Based on this simplistic model the gate errors would have to be lower than $\cO(10^{-5})$ in order to yield a fidelity above 0.9 using $\cO(10^4)$ gates. This gate precision is not within the scope of NISQ rotation and CNOT gates. If we instead assume that we have access to a fault-tolerant quantum computer with error correction we overcome this challenge but also lose access to arbitrary rotations which must now be synthesised using multiple T gates. We also incur further overhead in gate counts for the Hubbard model and long-range Ising model if the fault-tolerant quantum computer supports only local gates. This would be the case for the use of the surface code \cite{bravyi1998quantum, dennis2002topological}.   It is possible to take advantage of the parallel implementation of many identical rotations resulting in a reduction in the number of T gates required as detailed in \cite{bocharov15,gidney18}. Following these references, if we need to apply $N_R$ identical $R_z(\theta)$ rotations in parallel the number of T gates required is
\begin{equation}
    N_T = 4N_{R} + \log_2(N_{R})(1.15 \log_2(\frac{1}{E}\log_2(N_{\rm R}))+9.2),
\end{equation}

where $E$ is the the error per synthesised rotation. We set the requirement of introducing 1\% error across all rotation gates per time step sweep such that the error per synthesised rotation is $1\% / N_{\Sigma R}$ where $N_{\Sigma R}$ is the total number of rotation gates used in the simulation, as compared to $N_R$ which will be the number of identical rotations executed in parallel within a single time step sweep. The numbers for $N_{\Sigma R}$ for the Ising and Hubbard models appear in Tables \ref{tab:hubbard}, \ref{tab:ising} and \ref{tab:isingnn} under $R_z(\theta)$ gate counts for the Hubbard, long-range Ising and nearest neighbour Ising models respectfully. Hence we find the number of T gates required to construct the rotation gates for the full evolution using a fault-tolerant system for each model. For the Hubbard model this amounts to $4.6\times10^5$ T gates, and for the long-range Ising model, with the reduced ability to parallelise limiting the advantage of this rotation gate synthesis method, the resulting T gate count is $1.5\times10^7$. The most optimistic requirements are expected for the nearest neighbour Ising model where we benefit from the ability to execute many identical coupling terms in parallel as well as the absence of Jordan Wigner strings. This results in a T gate count for this model of $1.7\times 10^5$.

In such a fault-tolerant system the next step would be to calculate the resources required per T gate both in terms of logical and physical gates and qubits as a single T gate must be distilled from a larger set of noisy T gates \cite{litinski2019magic, bravyi2005universal}. These distillation processes are likely to dominate the resource requirements and set the limit on the logical gate fidelity required and the consequent physical qubit requirements.

In our analysis we determine the minimum hardware requirements in order for a digital quantum simulator to compete with experimentally realistic purpose built analogue devices at the task of simulating  the continuous time dynamics of Hubbard and spin models. We have found that in order to be able to simulate these on NISQ systems, would require on the order of at least $10^4$ gates, including this order of rotations. For a fault-tolerant system this many rotations would correspond to at least $10^5$ T gates. These are likely to dominate the resource requirements as each T gate is generated via an expensive distillation process.

\subsubsection{Post-Trotter simulation cost}
Recent post-Trotter quantum simulation algorithms based on Quantum
Signal Processing~\cite{Low2016HamSim} achieve optimal scaling with respect to time, error,
and system size like $\mathcal{O}\left(\alpha t+\log(1/\epsilon)\right)$,
but in terms of queries to unitaries $U=\ketbra{0\cdots0}{0\cdots0}\otimes H/\alpha+\cdots$
that block-encode the Hamiltonian up to some positive constant $\alpha$~\cite{Low2016Qubitization}.
The quantum gate cost is then dominated by this query cost multiplied by gate cost of synthesizing a controlled-block-encoding. The gate cost of other additional circuit
elements such as reflections by $\ketbra{0\cdots0}{0\cdots0}$ and
applying single-qubit rotations, are strongly sub-dominant to this block-encoding
and will be ignored.

Once the quantum gate cost of this block-encoding is accounted for,
the scaling of gate cost with respect to system size can be worse than Trotter methods.
For any Hamiltonian that is a sum of $K$ Pauli
operators $H=\sum_{j}\alpha_{j}P_{j}$ with positive coefficients, the block-encoding $U=\textsc{Prep}^{\dagger}\cdot\textsc{Select}\cdot\textsc{Prep}$
is generically constructed from the circuit preparing a quantum
state $\textsc{Prep}\ket{0\cdots0}=\frac{1}{\sqrt{\alpha}}\sum_{j}\sqrt{\alpha_{j}}\ket j\ket{\lambda}_{j}$, with a target state $\ket{\lambda}_{j}$
and a select circuit applying the Hamiltonian terms $\textsc{Select}=\sum_{j}\ketbra jj\otimes P_{j}$.
In the worst case, $\textsc{Prep}$ costs $\mathcal{O}\left(K\right)$
T gates, and controlled-$\textsc{Select}$ costs $\le4K$ T~\cite{Babbush2018encoding}
gates. However, symmetries such as the presence of identical coefficients
can greatly reduce cost. For instance, if there are $M'$ groups of
identical coefficients, the T cost of $\textsc{Prep}$ is only
$\mathcal{O}\left(M'\right)$ and may be ignored. Using the fact that $\textsc{Select}^{2}=I$,
in the relevant limit where $\alpha t\gg\log(1/\epsilon)$, the prefactor
in the query cost approaches $2\alpha t$~\cite{Low2016Qubitization}. 

The 2D Hubbard Hamiltonian in any qubit representation has at most $K=11M$ distinct Pauli terms, and the constant
$\alpha\approx M\left(4J+3U/4\right)$. Assuming that $J=U$, the
cost of simulation is $\approx352M^{2}Jt\left(19/16\right)$ T gates.
The case of $M=100$ and $Jt=10$ has a T cost of $\approx4\times10^{7}$,
which is significantly larger than the Trotter approach. Though further circuit optimizations beyond this generic construction could in principle reduce the prefactor, this is over-shadowed by the $M^2$ scaling. The long-ranged
Ising model contains significantly more terms, and can be expected a exhibit an even worse T cost compared to the Trotter approach.

\section{Conclusions}
\label{sec:conclusions}

We have analysed the propagation of errors for analogue quantum simulators, and the minimum gate counts required for digital quantum simulation, on two well-known problems of interest in physics and materials science. This identifies requirements for quantum simulation to be quantitatively reliable in regimes beyond what we can calculate classically with existing methods.

For analogue quantum simulators, our main result is that the calibration level of current experiments is already sufficient to obtain observable values with errors bounded at the 1\% percent level for 2D systems, where the same accuracy is not possible for any known classical algorithm. This comparison was carried out at the quantum advantage point, meaning that as analogue systems are scaled up, even more digital logic gates would be required to match the accuracy that can be achieved in the analogue experiments. This demonstrates clearly that analogue simulators are the best means to accurately compute observables in those classes of systems that can be realised (at this calibration level) directly in experiments, and confirms that existing experiments are already past the requirements for quantum advantage. 

We are confident that this type of error analysis can be directly compared with ongoing experiments, also with the addition of verification techniques \cite{Carrasco2021a,Eisert2020a}. In this respect, the assumptions that we made can be verified using Hamiltonian learning  \cite{Carrasco2021a,li2020hamiltonian, Bairey2019, Evans2019, Valenti2019, Wang2017} or Liouvillian learning \cite{Pastori2022} techniques.

On the side of digital quantum simulation, we note that these gate counts are a factor of $10^3$ improvement over previous estimates for the same protocol, \cite{Childs9456}. This improvements arises from a number of contributing factors; our extension of the model from 1D to 2D so that we need to propagate only until $tJ=10$ rather than 100, our choice of focusing on errors in observables, rather than basing errors on the state fidelities and in part due to application of gate saving techniques. Further reductions are unlikely as we have already reached time steps of order 1 and in general the cost is lower bounded by the product of the number of qubits, propagation time, and the cost of each local term.  We note that our analysis can be directly extended to other protocols, including recent optimisations of the digital algorithm \cite{clinton2020}, by considering directly the propagation of local errors. 
Potential trade-offs between gate errors and Trotter errors on NISQ machines, and other optimal or hardware-specific gate counts for fault-tolerant systems will be a very intersting area for follow-up research, as we look towards the roadmap for development of universal quantum simulators for models that cannot be directly realised on analogue platforms.

While the present analysis on the side of analogue quantum simulation relies on our ability to derive microscopic models for AMO platforms, quantum simulation is rapidly developing also with solid-state platforms. We expect that this error analysis will also be applicable to future generations of those platforms, especially when combined with Hamiltonian and/or Liouvillian learning to separately confirm the effective microscopic models being implemented.

\section{Acknowledgements}

This work was supported by the European Union's Horizon 2020 research and innovation program under grant agreement No. 817482 PASQuanS. Work at the University of Strathclyde was supported by the EPSRC through grants EP/P009565/1, EP/T001062/1 and EP/T005386/1, and by the AFOSR through grant number FA9550-18-1-0064.

All data underpinning this publication are openly available from the University of Strathclyde KnowledgeBase https://doi.org/10.15129/ab3e0b31-fcee-432e-995f-687a7446d045.

\appendix

\section{Limits of classical simulation}
\label{app:classical}

In this appendix, we illustrate in more detail the limitations of classical simulation for the problem choice considered. We show that the errors introduced in the dynamics upon limiting the precision of the classical algorithms are significantly larger than those arising from experimental error sources.

\begin{figure}[t!]
  \includegraphics[width=8.5cm]{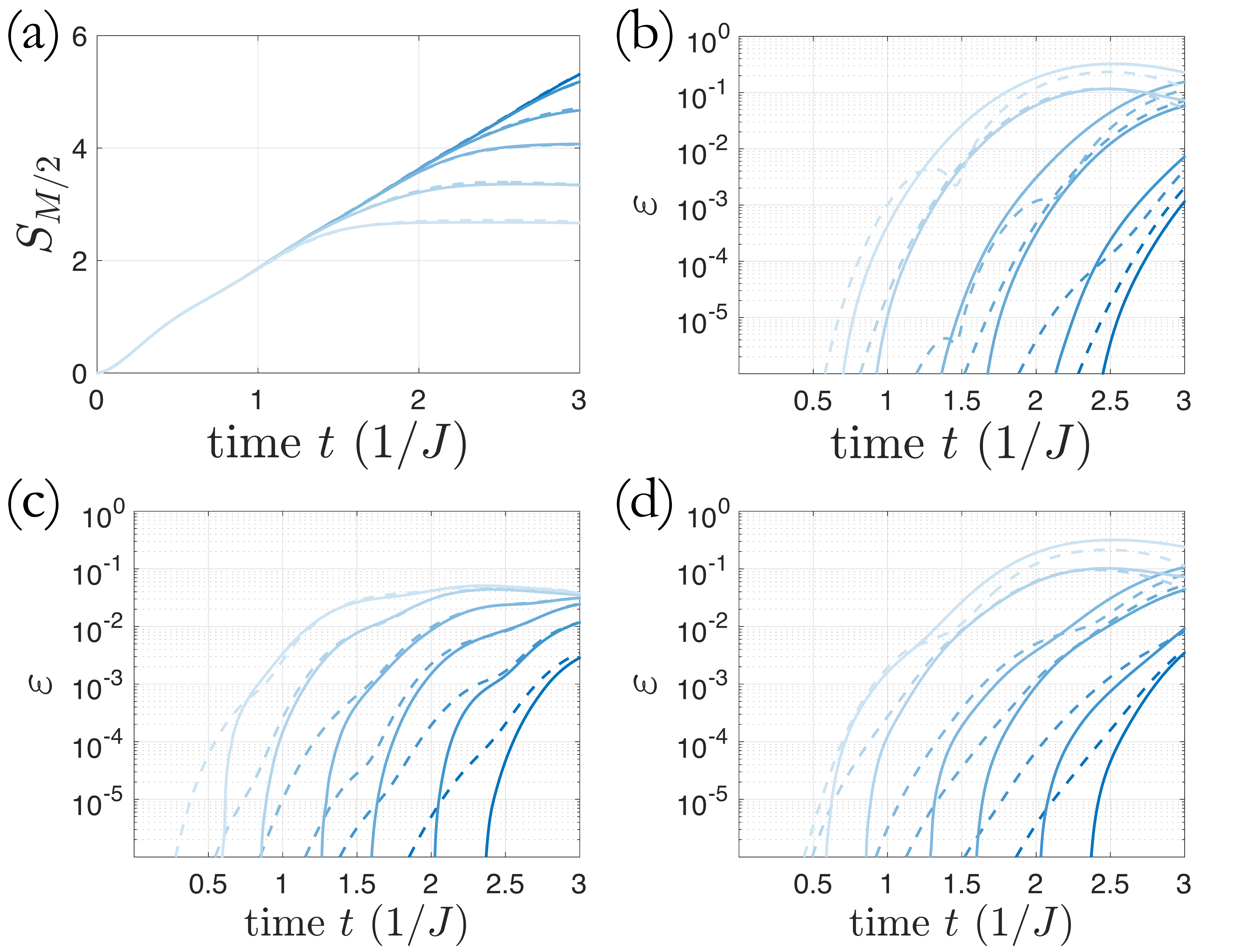}
  \centering
  \caption{Classical simulation with the single-site TDVP algorithm~\cite{PhysRevB.94.165116} (solid) and the TEBD algorithm~\cite{PhysRevLett.91.147902} (dashed) with a $4$th order left-right sweep Trotter decomposition. Results are shown for classical simulations with the bond dimensions, $D=\{512,256,128,64,32,16 \}$ (darkest to lightest), compared to $D=1024$. (a) Entanglement entropy at the centre bond. (b) Error (Eq.~\ref{equ_error}) in observable $O_n = c^{\dagger}_{n,\uparrow} c_{n,\uparrow} $. (c) $O_n = c^{\dagger}_{M/2,\uparrow} c_{n,\uparrow} $ (d) $O_n = c^{\dagger}_{M/2,\uparrow} c_{M/2,\uparrow} c^{\dagger}_{n,\uparrow} c_{n,\uparrow} $. System size, $M=20$, time-step $J\tau=0.001$ for TDVP and $J\tau=0.01$ for TEBD.}
  \label{SM_Fig1}
  \end{figure}

Throughout this analysis we consider the state-of-the art methods for propagating strongly-interacting systems, based around tensor networks. In particular, we have used tensor network techniques with time-evolving matrix product states (MPS), \cite{Daley_2004,PhysRevLett.93.076401,SCHOLLWOCK201196}. It has been shown that for these methods restricting the bond dimension limits the maximum entanglement entropy that can be described by the state~\cite{PhysRevLett.100.040501,Schuch:2008aa}. Since entanglement grows linearly after a global quench this restriction can quickly lead to large errors in the dynamics~\cite{Kliesch2014}. 

We demonstrate this in Fig.~\ref{SM_Fig1}(a) where we plot the time-dependence of the entanglement entropy of an initial product state, $|\Psi(0)\rangle = | \uparrow,\downarrow,\uparrow,\downarrow,\cdots\rangle$ after a quench. Here we use a two-site time-evolving block decimation (TEBD) algorithm~\cite{PhysRevLett.91.147902} with  particle number and total spin conservation. We use the  $4$th order Trotter decomposition considered in the main text. Note that one could perform this simulation using different MPS time evolution algorithms and so in the figure we perform the same analysis using the single-site time-dependent variational principle (TDVP)~\cite{PhysRevB.94.165116} (solid) and compare to the TEBD algorithm (dashed) where we can see similar behaviour with only small quantitative corrections. Note that for this TEBD algorithm with the  $4$th order Trotter decomposition we have sufficiently converged results for a time-step of $J\tau=0.01$, but for the TDVP algorithm, which has an error scaling of $\tau^3$~\cite{PhysRevB.94.165116}, we use a time-step of $J\tau=0.001$ to ensure that errors arising from discretizing time are negligible.

As the bond dimension required to capture a linear growth of entanglement must grow exponentially we can see in Fig.~\ref{SM_Fig1}(a) that truncating the bond dimension at each time step significantly limits the entanglement growth and causes distortion to the values of other observables. The bond dimension is truncated in the TEBD algorithm by performing singular value decompositions on each neighbouring pair of two-site tensor and then only retain components with the $D$ largest singular values which ensures the retention of the most important information at each step within the restrictions imposed. We have utilised the ITensor Library\cite{ITensor}, optimising performance by conserving quantum numbers.

At this point we can ask the central question; if we restrict the bond dimension to values that are efficiently simulable on classical computers, how do the resulting errors in the observables compare to the errors produced through experimental imperfections in analogue simulation? We include the results of this analysis on the Hubbard model (Eq.~\ref{Hubb_Mod}) with $U=J$ in Fig.~\ref{SM_Fig1}(b-d) for different observables where we can see that these errors grow exponentially in time and reach values that are significantly larger than those from a calibration uncertainty in analogue simulation (see Fig.~\ref{Fig2}), even reaching close to order 1. For most observables and only small truncation errors (i.e. bond dimension close to the exact case), the TDVP algorithm seems to have smaller errors initially, but these grow rapidly in time, quickly reaching the same values as the TEBD approach for the final times considered here. For lower bond dimension where there is a much larger truncation error, this difference between the two algorithms is not as pronounced. This analysis indicates that the timescales that we have considered in the main text are significantly beyond what is possible to simulate with these state of the art classical algorithms. 

There are additional algorithms for time-evolution~\cite{Paeckel:2019aa} and even different ways that we could perform the compression to a lower bond dimension such as through a variational approach~\cite{SCHOLLWOCK201196}. These will lead to quantitative differences, but we do not expect any to give a significant advantage or allow a classical simulation to evolve to significantly longer times as all algorithms ultimately rely on truncating the maximum allowed bond dimension in a similar way as the two algorithms explicitly considered. These two algorithms considered here were chosen as they perform the optimisation of the MPS in qualitatively different ways. The TEBD algorithm is conceptually simpler, applying two-site operations and then compressing the tensors by applying an SVD on each long-range. Whereas in the TDVP approach we optimise a single tensor variationally, while taking into account the effects of the entire Hamiltonian and as there is no compression of the bond dimension, in principle this algorithm finds the optimal global approximation of $|\psi(t+\tau) \rangle$ with the given bond dimension. However, the TEBD algorithm gives much more flexibility in the particular decomposition of the full time-evolution operator into the two-site gate sequence and in particular when applying these algorithms to this problem, we find that TEBD can give the same accuracy for a much larger time step. 

\section{Analysis of calibration errors}
\label{app:calibration}

In this section we provide additional details for the analysis on the effects of a calibration uncertainty on the dynamical behaviour of local observables and correlations. In particular, show that these errors do not depend on the system size (as long as boundary effects are unimportant) or even the initial state used and we discuss and justify an extrapolation of this behaviour to quantum simulation in higher dimensions.

When introducing callibration errors, for each model we sample the parameters globally (i.e. the model parameters are always homogeneous across the system) from a Gaussian distribution, with standard deviation given by the calibration error $\Delta$, and then compare to the results of a simulation with parameter values at the mean of this distribution. 

\begin{figure}[t]
  \includegraphics[width=8.5cm]{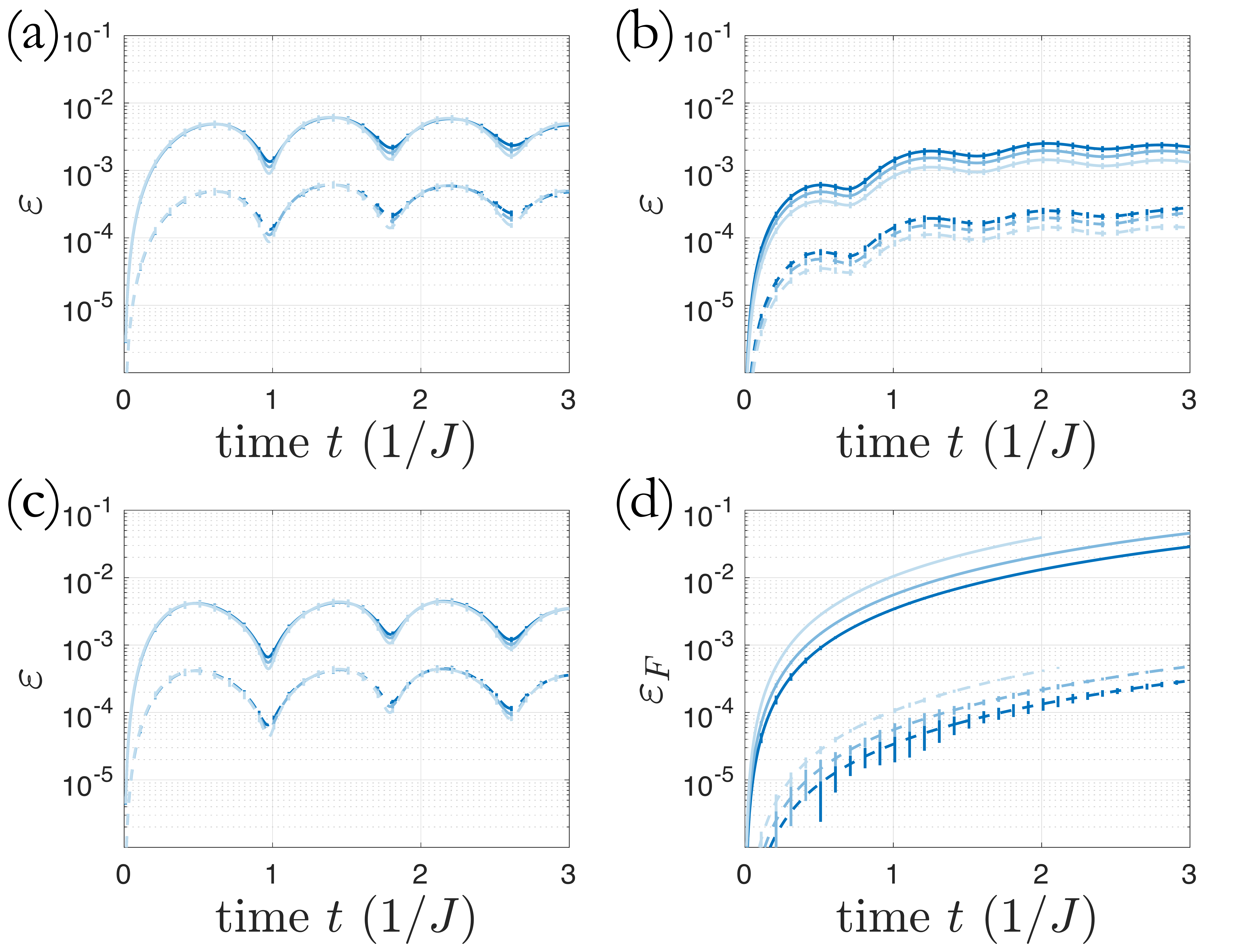}
  \centering
  \caption{Errors in the observables of a Hubbard model due to a  calibration uncertainty of the parameters for system sizes $M=\{20,32,60 \}$ (Darkest to lightest). We compare calibration errors $\Delta = 1\%$  (solid) and $\Delta = 0.1\%$ (dashed). (a) Error (Eq.~\ref{equ_error}) in $O_n = c^{\dagger}_{n,\uparrow} c_{n,\uparrow} $. (b) $O_n = c^{\dagger}_{M/2,\uparrow} c_{n,\uparrow} $. (c) $O_n = c^{\dagger}_{M/2,\uparrow} c_{M/2,\uparrow} c^{\dagger}_{n,\uparrow} c_{n,\uparrow} $. (d) Error in the state fidelity (Eq.~\ref{equ_error_fid}).}
  \label{SM_Fig2}
  \end{figure}

\subsection{System size dependence}

In Fig.~\ref{SM_Fig2} we show the results for the Hubbard model, $U=J=1$, beginning in the product state, $|\Psi(0)\rangle = | \uparrow,\downarrow,\uparrow,\downarrow,\cdots\rangle$.  It is apparent that there is only a very weak system size dependence on the errors in the observables we consider, where the small variations arise from boundary effects. This indicates that, for timescales we consider here, which are much smaller than the time it takes for information to spread from the centre to the boundary, we can reliably extrapolate this error behaviour to larger systems, such as those that can be realised in cold atom experiments.

Additionally, we compare the errors in the observables to errors in the total state fidelities defined by,
\begin{equation}\label{equ_error_fid}
\varepsilon_F = 1 - |\langle \psi_{\rm sim}(t)| \psi_{\rm ex}(t) \rangle|^2,
\end{equation}
where we see that, as expected, the errors in the chosen observables are smaller than the errors in the entire wavefunction. We can also see that while the observable errors do not depend on system size, the errors in the fidelities grow as the system size is increased. This show the importance of choosing how to quantify errors and success rates for a quantum simulation based on the problem choice at hand.

\begin{figure}[t]
  \includegraphics[width=8.5cm]{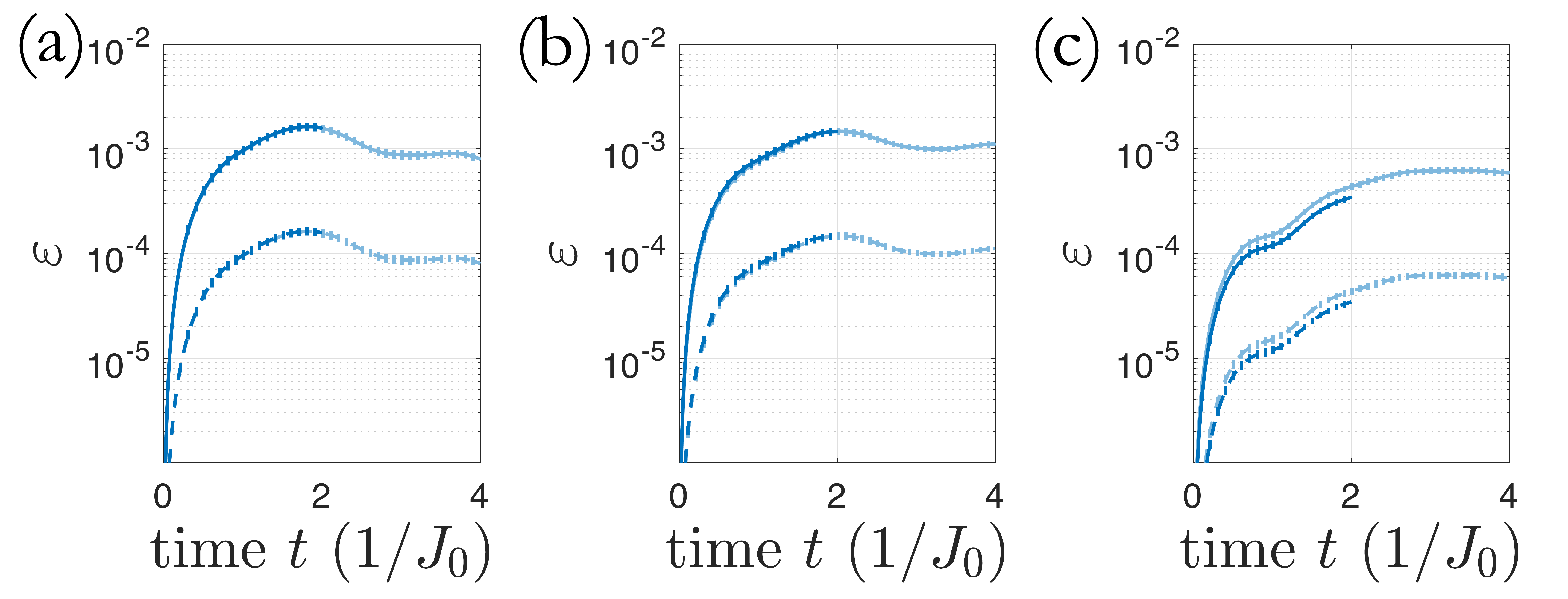}
  \centering
  \caption{Errors in the observables of a Transverse Ising model with algebraically decaying interactions, $\alpha=2$, due to a global calibration uncertainty of the parameters for system sizes $M=\{20,32 \}$ (light blue, dark blue). Note that the $M=32$ simulation stops at $tJ_0 = 2$ due to errors arising from truncating the bond dimension. We compare calibration errors $\Delta = 1\%$ (solid) and $\Delta = 0.1\%$ (dashed). (a) Error (Eq.~\ref{equ_error}) in $O_n = s_x^{n}$. (b) $O_n = s_+^{M/2} s_-^{n}  $. (c) $O_n = s_z^{M/2} s_z^{n} $.}
  \label{SM_Fig3}
  \end{figure}

\begin{figure}[t]
  \includegraphics[width=8.5cm]{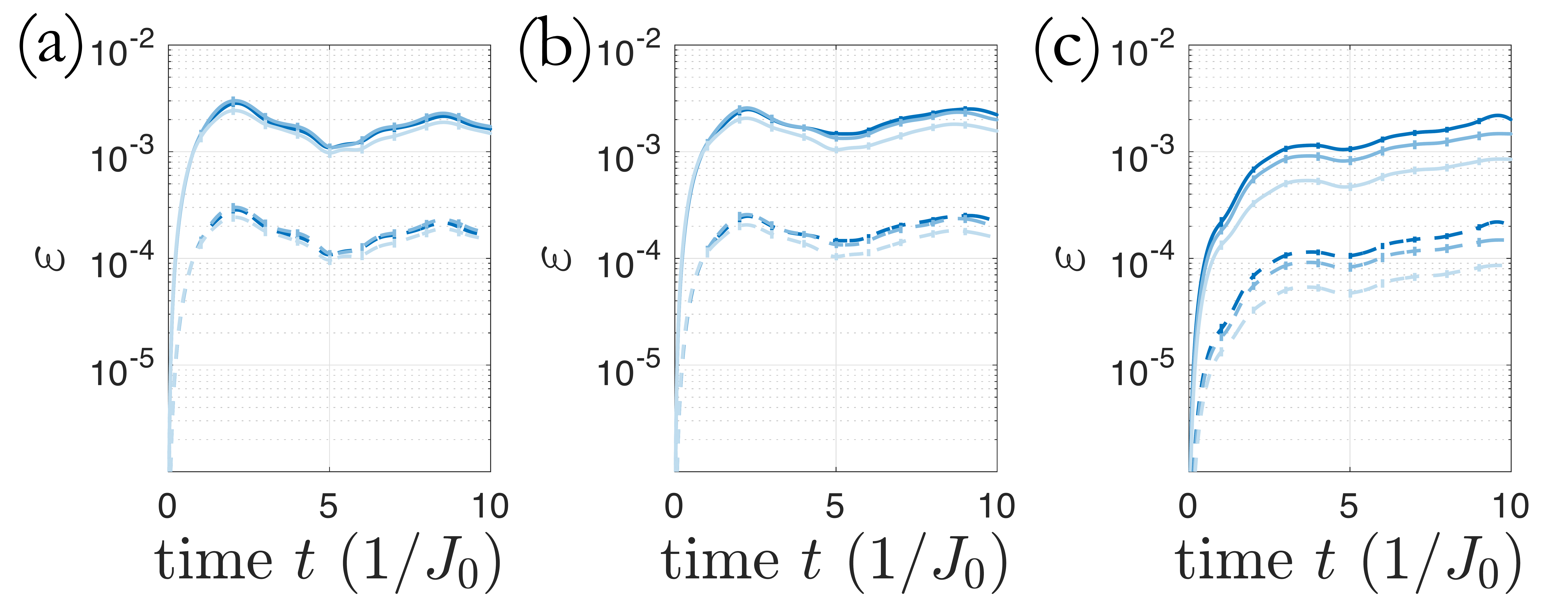}
  \centering
  \caption{Errors in the observables of a Transverse Ising model with nearest neighbour interactions, $\alpha \rightarrow \infty$, due to a calibration uncertainty of the parameters for system sizes $M=\{20,32,60 \}$ (Darkest to lightest). We compare a calibration errors $\Delta = 1\%$ (solid) to $\Delta = 0.1\%$ (dashed). (a) Error (Eq.~\ref{equ_error}) in $O_n = s_x^{n}$. (b) $O_n = s_+^{M/2} s_-^{n}  $. (c) $O_n = s_z^{M/2} s_z^{n} $.}
  \label{SM_Fig4}
  \end{figure}

We also consider the transverse Ising model, $J_{0}=1=B=1$, beginning with an initial state, $|\Psi(0)\rangle = | +,+,+,+\cdots\rangle$, and in Fig.~\ref{SM_Fig3} we show the results for the case for algebraically decaying interactions, $J^{(n,m)} = J_0/|n-m|^\alpha$ with $\alpha=2$ and in Fig.~\ref{SM_Fig4} we show the results for a long-range interaction, $\alpha \rightarrow \infty$. We can see that in both cases the errors in the observables only depend on the system size very weakly.

\subsection{Dependence on the initial state}

\begin{figure}[t]
  \includegraphics[width=8.5cm]{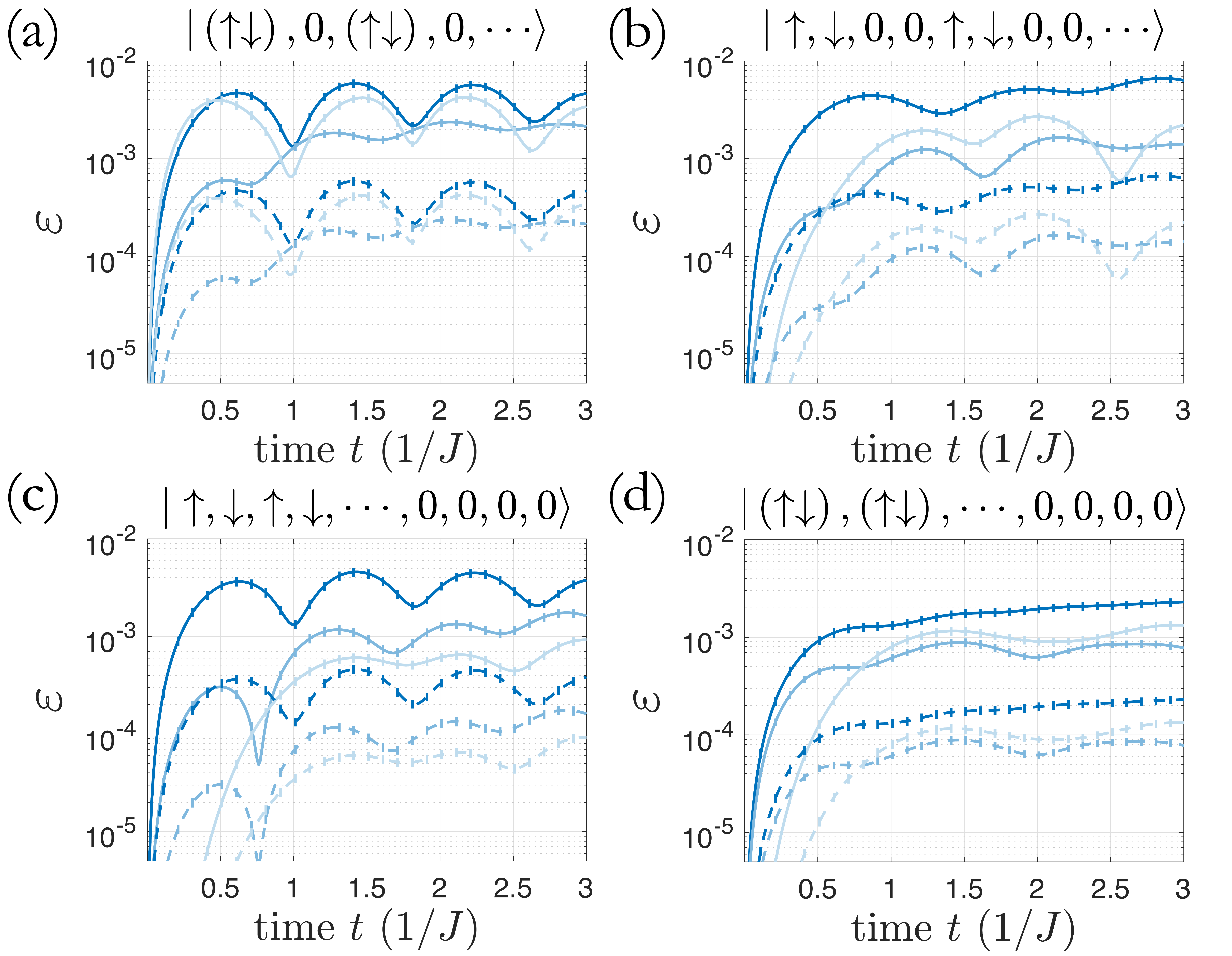}
  \centering
  \caption{Comparing errors generated due to a calibration uncertainty for different initial states in the Hubbard model. The errors (Eq.~\ref{equ_error}) for (Darkest to lightest) $O_n = c^{\dagger}_{n,\uparrow} c_{n,\uparrow} $, $O_n = c^{\dagger}_{M/2,\uparrow} c_{n,\uparrow} $ and $O_n = c^{\dagger}_{M/2,\uparrow} c_{M/2,\uparrow} c^{\dagger}_{n,\uparrow} c_{n,\uparrow} $. We compare a $1\%$ calibration error (solid) and a $0.1\%$ calibration error (dashed). (a-b) With periodic initial states. (c-d) Initial states that are empty on the right half. For all we use a system size of $M=20$.}
  \label{SM_Fig7}
  \end{figure}

In order to determine whether the observed behaviour of the errors is due to some special feature of our initial state we compare the errors due to a calibration error in the Hubbard model when beginning in different initial product states. In Fig.~\ref{SM_Fig7} we plot the results for different observables for both $1\%$ and $0.1\%$ calibration error. In  Fig.~\ref{SM_Fig7}(a-b) we use states that are periodic throughout the entire system and in Fig.~\ref{SM_Fig7}(c-d) we use initial states that are filled (and periodic) on the left half but empty on the right. While there is different oscillatory behaviour in each of these simulations the qualitative dependence is very similar, with a rapid initial increase in error followed by a much more gradual growth. We can see that all initial states considered saturate at similar values, leading us to believe that our findings are general for initial product states.

The analysis in this and previous sections greatly indicate that the scaling of errors in observables due to a calibration uncertainty are not significantly affected by either the initial state, the system size or even the model considered. By considering these errors and the problem of time-evolution, we thus have a very general way of benchmarking quantum simulation. 

\subsection{Extrapolation to two-dimensions}
\label{app:Extrap}

We have shown that we are able to reliably extrapolate our results to larger 1D systems as well as simulations that use different initial states. Now in this section we justify our further extrapolation of the effect of a calibration error to the quantum simulation of two-dimensional (2D) Hubbard models. 

\begin{figure}[t]
  \includegraphics[width=8.5cm]{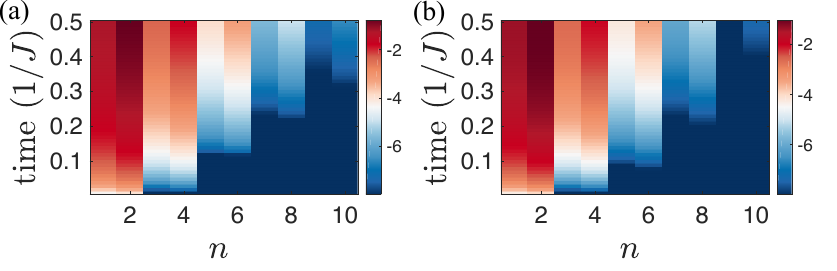}
  \centering
  \caption{Comparison of the spreading of the off-diagonal correlations, $C_n = \langle c^{\dagger}_{M/2,\uparrow} c_{n,\uparrow} \rangle$, in time for a 1D Hubbard chain (a) and one leg of a two-leg ladder (b). In both cases we begin in an anti-ferromagnetic state $|\uparrow,\downarrow,\uparrow,\downarrow,\cdots \rangle$, where nearest neighbour sites always have opposite spin. Colours on a log (base 10) scale.}
  \label{SM_Fig6}
  \end{figure}

The Lieb-Robinson bounds limit the distance which correlations can propagate for locally interacting Hamiltonians, restricting the transport of information~\cite{Poulin2015,Kliesch2014,childs2019trotter}. As a result we do not expect any fundamental changes to occur when going from 1D to 2D for these types of simulations of dynamics after a global quench. Although the quantitative behaviour can be affected, importantly we expect the scaling of these errors in time to be qualitatively the same. In Fig.~\ref{SM_Fig6} we quantitatively compare the spreading in the off-diagonal correlations in the 1D chain (Fig.~\ref{SM_Fig6}(a)) and the two-leg ladder (Fig.~\ref{SM_Fig6}(b)). It is clear that these spread at the same rate, validating our assumption that our analysis so far of information spreading in these types of systems is valid for higher dimensions. 

\begin{figure}[t]
  \includegraphics[width=8.5cm]{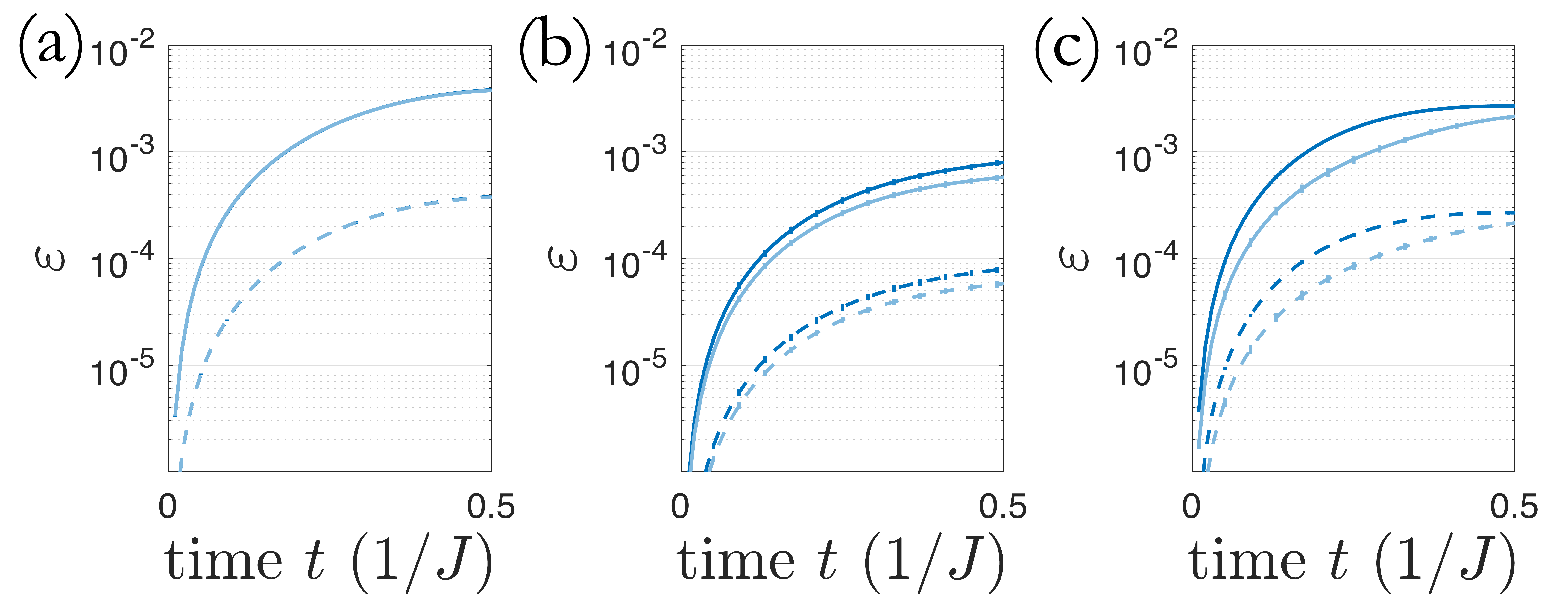}
  \centering
  \caption{Errors in the observables of a Hubbard model on a two-leg ladder due to a calibration uncertainty of the parameters for system sizes $M=\{20,40 \}$ (dark blue, light blue). We compare a $1\%$ calibration error (solid) and a $0.1\%$ calibration error (dashed). (a) Error (Eq.~\ref{equ_error}) in $O_n = c^{\dagger}_{n,\uparrow} c_{n,\uparrow} $. (b) $O_n = c^{\dagger}_{M/2,\uparrow} c_{n,\uparrow} $. (c) $O_n = c^{\dagger}_{M/2,\uparrow} c_{M/2,\uparrow} c^{\dagger}_{n,\uparrow} c_{n,\uparrow} $.}
  \label{SM_Fig5}
  \end{figure}

We further verify that the errors in question have similar behaviour to the correlations for the two-leg ladder by repeating our analysis of errors in observables caused by miscalibration for this new structure. In Fig.~\ref{SM_Fig5} we show the growth of these errors produced through calibration uncertainty in the parameters $U$ and $J$. Although we are limited to shorter times due to an increased growth of entanglement causing the classical algorithm to become more numerically demanding, we can see that the errors appear to saturate at similar values and have the same scaling behaviour in time. This indicates that there is no fundamental change upon including dynamics in a second dimension. 

Experimental proof would would be optimal in order to fully verify this behaviour but even without it we can be confident that the scaling of errors in these observables due to a calibration uncertainty is significantly smaller than that due to limiting the precision of the classical algorithm (see Fig.~\ref{SM_Fig1}). Having repeated our analysis for different initial states, structures, calibration errors and system sizes is clear that while the scaling of errors due to experimental sources may have quantitative differences the overall scaling will still be extremely favourable compared to the classical algorithms, which will perform even more poorly in 2D.

\section{Additional sources of errors in analogue simulators}
\label{app:othererrors}

In the main text we explicitly compared the errors due to a calibration uncertainty to heating due to laser fluctuations and decoherence effects. However, there are a number of other potential sources of errors in an experimental realisation of analogue simulation with ultra-cold atoms. In this appendix we explicitly consider an error in the preparation of the initial state and then separately errors due to the presence of unwanted terms in the physical Hamiltonian of the simulator. Through our quantitative analysis of the dynamical errors in observables we argue that for the situations and timescales that we consider, that these give errors orders of magnitude smaller, compared to those arising from a calibration uncertainty and so can be neglected in our error budget analysis discussed in the main text. 

\begin{figure}[t]
  \includegraphics[width=8.5cm]{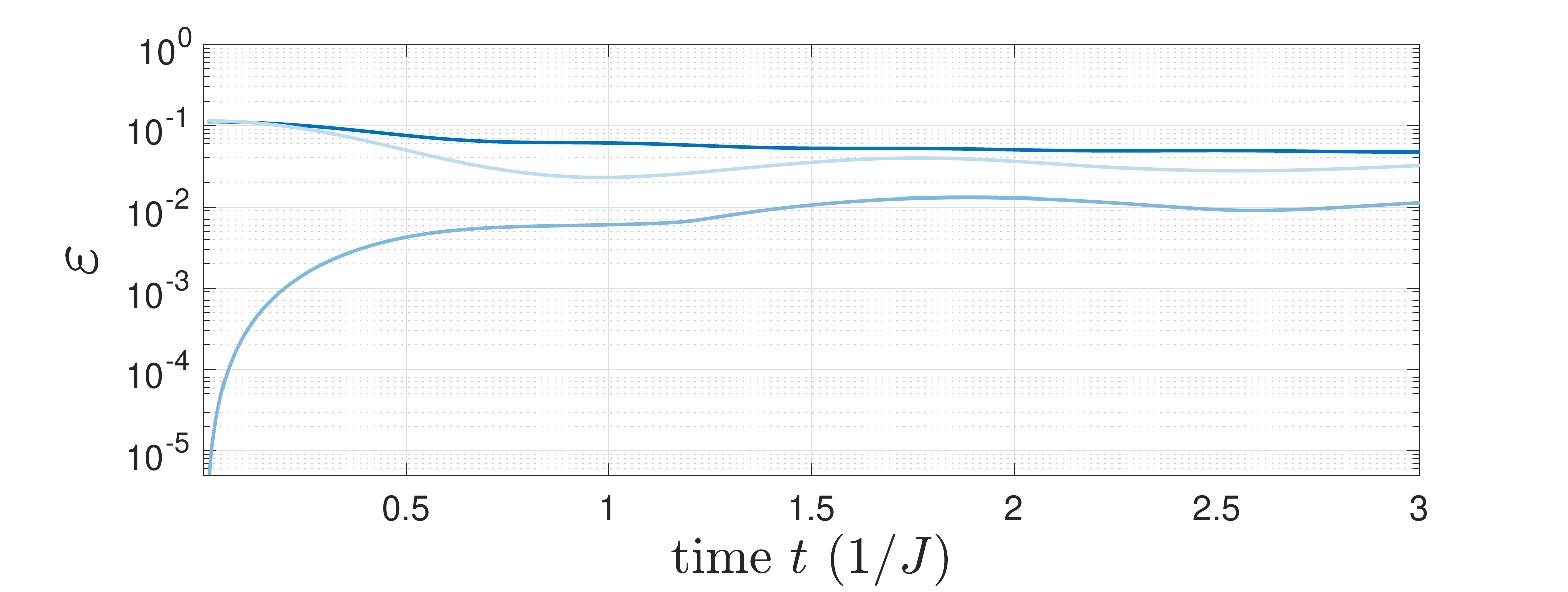}
  \centering
  \caption{Errors in the Hubbard model due to an error in the preparation of the initial state for system size, $M=20$. The error (Eq.~\ref{equ_error}) averaged over removing a single particle at one lattice site in the initial state. Errors in (Darkest to lightest) $O_n = c^{\dagger}_{n,\uparrow} c_{n,\uparrow} $, $O_n = c^{\dagger}_{M/2,\uparrow} c_{n,\uparrow} $ and $O_n = c^{\dagger}_{M/2,\uparrow} c_{M/2,\uparrow} c^{\dagger}_{n,\uparrow} c_{n,\uparrow} $}
  \label{SM_Fig8}
  \end{figure}

\subsection{State preparation}

In Fig.~\ref{SM_Fig8} we plot the errors in the observables of the Hubbard model due to an error in the preparation procedure of the initial state, $|\Psi(0)\rangle = | \uparrow,\downarrow,\uparrow,\downarrow,\cdots\rangle$. In order to model this we remove a fermion from a single lattice site at time $t=0$ and perform different simulations for the removal at different sites. In Fig.~\ref{SM_Fig8} we average the global error (Eq.~\ref{equ_error}) over removing the fermion at each lattice site. While the average error is large, even at initial times, this is dominated by local errors around the position of the incorrectly prepared site and at positions further away from the empty site then the errors are negligible for the timescales considered. 

Note that these are the error values, given that the state has been incorrectly prepared. But in state-of-the-art simulations, this will typically only occur $\sim1\%$ of the time~\cite{Mazurenko2017}. So these values should be multiplied by $10^{-2}$ before comparing to the errors induced through the calibration errors. This particular error could be further reduced by performing post selection of the simulations where if it is clear that an initial error has occurred, then these simulations can be neglected.

\subsection{Additional terms in the Hamiltonian}

Another potential source of error is additional mechanisms influencing the dynamics in analogue simulation experiments. These processes will most likely have a much smaller magnitude compared to the onsite and nearest neighbour tunnelling terms but could lead to large errors at long times. The parameters in the Hubbard model which best describes the dynamics of trapped atoms, can be derived quantitatively from first principles~\cite{PhysRevLett.81.3108}, and while first order processes can be made to be negligible in the experiment by carefully controlling the laser parameters, second-order terms that can arise from strong interactions could play a role if the regimes probed are not chosen carefully.

The most likely additional terms take the form of next nearest neighbour tunnelling,
\begin{equation} \label{Jnn_equ}
H_{Jnn} = -J_{nn} \sum_{n,\sigma} \left( \hat{c}^{\dagger}_{n,\sigma} \hat{c}_{n+2,\sigma} + h.c. \right),
\end{equation}
or nearest neighbour interaction,
\begin{equation} \label{Vnn_equ}
H_{Vnn} = V_{nn} \sum_n \left( \hat{n}_{n,\uparrow} + \hat{n}_{n,\downarrow} \right) \left( \hat{n}_{n+1,\uparrow} + \hat{n}_{n+1,\downarrow}\right).
\end{equation}

\begin{figure}[ht]
  \includegraphics[width=8.5cm]{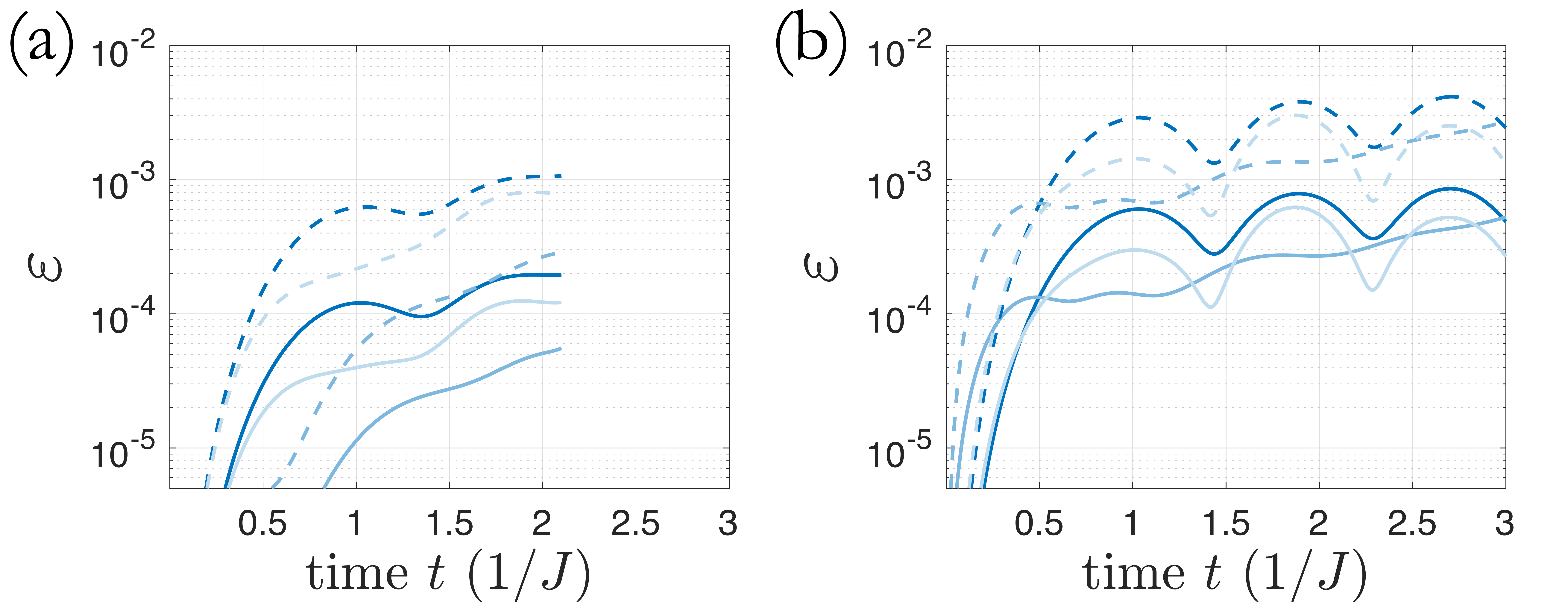}
  \centering
  \caption{Errors in the Hubbard model due to  the presence of additional terms in the Hamiltonian for system size, $M=20$. Errors in (darkest to lightest) $O_n = c^{\dagger}_{n,\uparrow} c_{n,\uparrow} $, $O_n = c^{\dagger}_{M/2,\uparrow} c_{n,\uparrow} $ and $O_n = c^{\dagger}_{M/2,\uparrow} c_{M/2,\uparrow} c^{\dagger}_{n,\uparrow} c_{n,\uparrow} $. We plot the cases where the additional terms have a magnitude of $1\%$ (solid) and $5\%$ (dashed) compared to the nearest neighbour tunnelling $J$ and the onsite interactions $U$. (a) Errors due to including a next-nearest neighbour tunnelling term, Eq.~\ref{Jnn_equ} (note this value stops at $tJ=2.1$ due to errors arising from truncating the bond dimension). (b) Errors due to including a nearest neighbour interaction, Eq.~\ref{Vnn_equ}.}
  \label{SM_Fig9}
  \end{figure}

In Fig.~\ref{SM_Fig9} we plot the resulting errors in the observables upon including these additional terms at the level of $5\%$ and $1\%$. For values around $1\%$ we can see that the errors are smaller than those induced from a calibration error on the level of $0.1\%$. This indicates that for the timescales that we are considering, the errors from a calibration error is much more significant than those arising from these types of Hamiltonian correction terms.

\section{Trotter decompositions}
\label{app:trotter}

In this appendix we present our estimates for the critical time steps at which point a digital simulator will have an error in a particular observable comparable to the errors produced from a calibration uncertainty in an analogue simulator. 

To implement these types of dynamical simulations on a digital quantum simulator, it is necessary to discretise the evolution into time steps using the Trotter decomposition and then further decompose these evolution operations into the logic operations (gates) that are native to the digital quantum simulator. This is similar to performing the simulations on a classical computer, the difference being that with current (and near term) architectures for quantum simulation we are limited in the total number gates that we can perform accurately. This limitation is due to the finite coherence times of these systems, which limits the fidelity of the gates used to implement the simulation and hence may result in an error that increases with each gate. In order to analyse the feasibility of executing these simulations on a digital quantum computer it is thus critical to understand how many gates will be necessary to execute a simulation which has comparable errors to the analogue quantum simulation. In order to calculate this we must first investigate the behaviour of errors in observables that arise from executing the Trotter decompostion in order to digitize time in the simulation. We investigate how these errors behave with increasing the time step, $\tau$, in order to find the largest time step possible at which we can achieve similar precision in a digital simulation to an analogue one. Although a larger $\tau$ introduces more error it allows for a smaller number of time steps which is desirable in order to minimize the number of gates required. As such, balancing the time decomposition error with the total number of required logic gates is crucial. 

As described in the main text, we use the error values caused by calibration error in our classical simulations of an analogue simulator as an \textit{error budget} and we attempt to find the critical time step, $\tau^*$, where the errors that arise from a discretization in time match this budget. This depends on the particular way the discretization of the time-evolution operator is performed (the Trotter decomposition) and so we compare to many of the most standard forms that are routinely used for performing MPS time-evolution simulations on a classical computer.

Once we have calculated the number of gates required for a digital quantum simulator to execute a simulation with comparable errors to an analogue one we can then ask how accurate these gates need to be and assess the feasibility of achieving this in the near term. An alternative method would be to include gate error from the start when trying to find a decomposition which results in errors matching the error budget but there is no assurance that this would be possible and so we instead choose to only consider Trotter error when finding a decomposition and estimating the required resources.

\subsection{Hubbard Model}

First we present the results for the Hubbard model. As this model contains only onsite and nearest neighbour terms it is particularly suited for a decomposition into single and two site gates of which there are a variety of different approaches. 

If we split the Hamiltonian $H$ up into a sum of terms which act on nearest neighbour sites $H = \sum_n^{M-1} H_{n,n+1}$, then we can write the time-evolution operator which evolves the system for a single time step $\tilde{U}(\tau)$ as
\begin{equation}
\begin{split}
\tilde{U}(\tau) & = e^{-i H \tau} \\
&\approx \prod_n^{M-1} e^{-i H_{n,n+1} \tau},
\end{split}
\end{equation}
where the second term is only approximately equal to due to the fact that $H_{n,n+1}$ does not commute with nearest neighbouring term. The above equation would be the simplest first order Trotter decomposition with an error that scales with $\tau$, but there are many higher order approaches which work in a similar way.

Here we have considered two such approaches, first, a decomposition involving the successive sweeping left and then right through the 1D chain~\cite{PhysRevA.60.1956} and second, a successive application on even and then odd bonds~\cite{SCHOLLWOCK201196}. In order to define the decompositions for the former, we introduce the notation,
\begin{equation}
(s) = \prod_{n=1}^{M-1}  e^{-i H_{n,n+1} s\tau/2} = \prod_{n=1}^{M-1}\tilde{U}_{n,n+1}(s\tau/2),
\end{equation}
which indicates a single left-right sweep through the system with a time-step $s\tau/2$. This means that we first apply $\tilde{U}_{1,2}(s\tau/2)$, then $\tilde{U}_{2,3}(s\tau/2)$ and so on until $\tilde{U}_{M-1,M}(s\tau/2)$. Additionally we define the reverse operation, corresponding to a similar sweep but from the right to the left,
\begin{equation}
(s)^T = \prod_{n=M-1}^{1}  e^{-i H_{n,n+1} s\tau/2} = \prod_{n=M-1}^{1}\tilde{U}_{n,n+1}(s\tau/2).
\end{equation}

This allows us to define the second order sweep decomposition as,
\begin{equation} \label{equ_2nd_decomp_LR}
\tilde{U}_{\rm sweep}(\tau) = (1)(1)^T,
\end{equation}
which has an error that scales with $\tau^2$.

\begin{figure}[t]
  \includegraphics[width=8.5cm]{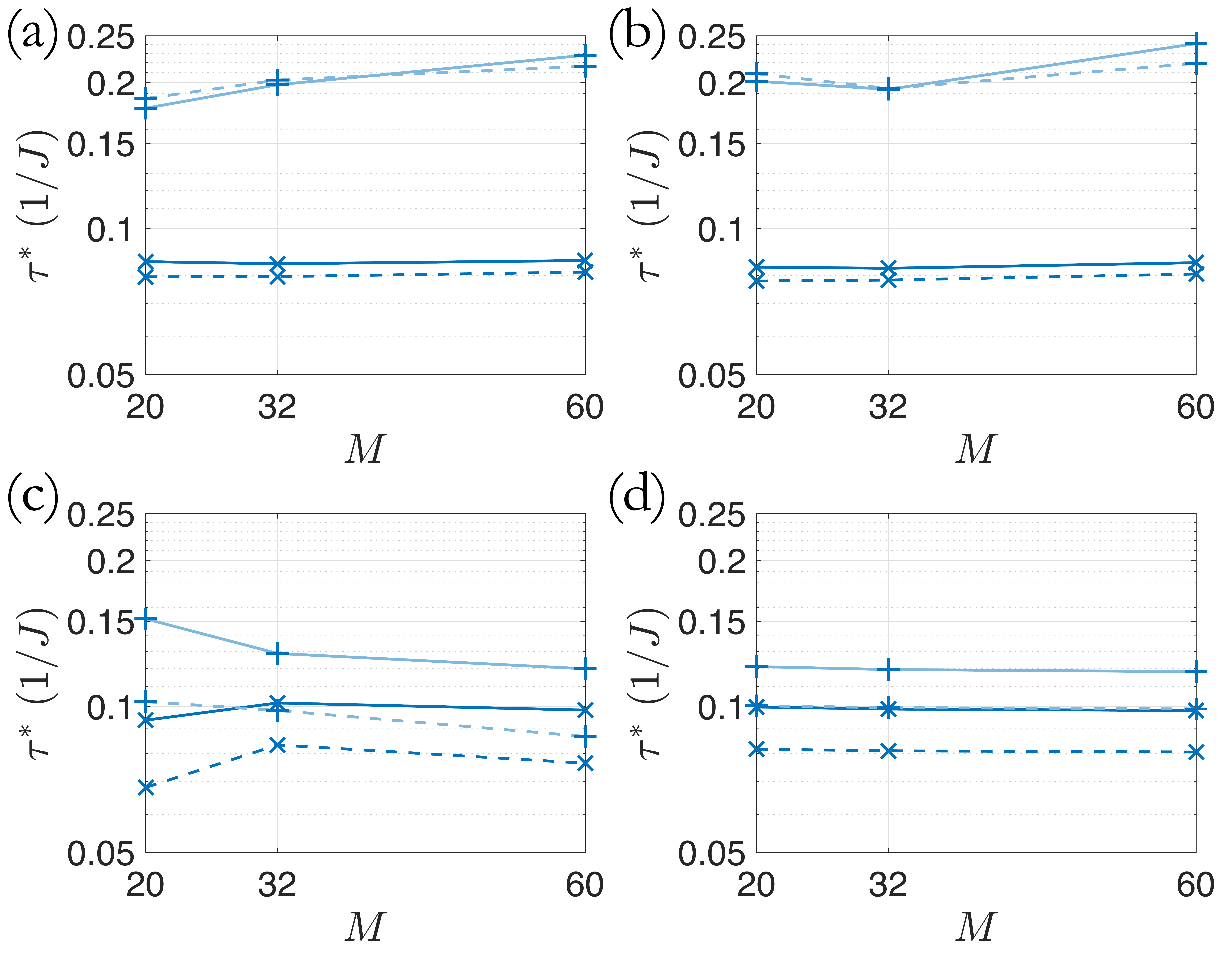}
  \centering
  \caption{Comparison of critical time steps using different methods of second order Totter decompositions. Shown are the critical time steps, $J\tau^*$, at which a digital simulation executed by the either method would result in Trotter errors of equal magnitude to those in an analogue simulation due to a calibration error of $0.1\%$ at $tJ = 3$. We compare a left-right sweep decomposition (crosses) defined in Eq.~\ref{equ_2nd_decomp_LR} to an odd-even split (plus signs) defined in Eq.~\ref{equ_2nd_decomp_OE}. Additionally, we compare the cases where each two site gate contains both the tunnelling and interaction terms (solid) to the case where these pieces are also split into separate time-evolution operations (dashed). (a) Errors in $O_n = c^{\dagger}_{n,\uparrow} c_{n,\uparrow} $. (b) $O_n = c^{\dagger}_{M/2,\uparrow} c_{n,\uparrow} $. (c) $O_n = c^{\dagger}_{M/2,\uparrow} c_{M/2,\uparrow} c^{\dagger}_{n,\uparrow} c_{n,\uparrow} $. (d) Error in the state fidelity (Eq.~\ref{equ_error_fid}).}
  \label{SM_Fig10}
  \end{figure}
  
We could alternately execute a second order Suzuki-Trotter decomposition by applying the evolution operators of all the odd bonds followed by those of the even bonds rather than sweeping from left to right and back. This is defined by
\begin{equation} \label{equ_2nd_decomp_OE}
\tilde{U}_{\rm OE}(\tau)  = e^{-i H_{\rm odd} \tau/2} e^{-i H_{\rm even} \tau} e^{-i H_{\rm odd} \tau/2}.
\end{equation}
This involves a single application of the time-evolution operator for all odd bonds with half a time step, followed by the application on all even bonds with a full time-step which is finally followed up by evolving the odd bonds again for another half time-step. This works particularly well for the Hubbard model as it only contains onsite and nearest neighbour terms so that all $H_{\rm even}$ commute with one another and all $H_{\rm odd}$ commute with one another. This means that we can split up the time-evolution operator for the odd (and even) bonds into a product of time-evolution on each bond separately with no error. 

In Fig.~\ref{SM_Fig10} we perform the same simulation as considered in the main text, where we begin in an initial product state, $|\Psi(0)\rangle = | \uparrow,\downarrow,\uparrow,\downarrow,\cdots\rangle$, and then evolve with the Hubbard Hamiltonian with $U=J$. Here we compare the errors in observables when using the two defined second order trotter decompositions, Eq.~\ref{equ_2nd_decomp_LR} and \ref{equ_2nd_decomp_OE}, in order to simulate the time-evolution. By performing a fit of the errors at $tJ=3$, we extrapolate the critical time-step $\tau^*$ at which the errors in each observable due to the Trotter decomposition and finite time step become equal to those generated by a particular level of calibration error. 

We also investigate the difference between treating the two site terms in the Hamiltonian with and without splitting up the tunnelling and interaction terms. Splitting up these terms typically leads to a larger decomposition error, but each piece can be approximated with fewer logic gates in an experimental realisation. We find that in terms of this error budget analysis that this consideration leads to qualitatively similar values. 

For these simulations, we find that the odd-even decomposition permits larger  critical time step values which do not strongly depend on system size or on which observable we use to determine them.



Higher order decompositions offer better scaling of error with time step size which motivates us to carry out a similar analysis on a variety of different fourth order decompositions. There are more options for how to carry out higher order decompositions, here we compare the different sweep decompositions derived in Ref.~\cite{PhysRevA.60.1956}, which for clarity we write here
\begin{equation} \label{equ_4th_decomp_LR}
\begin{split}
\tilde{A}_{\rm sweep} & = (1)^T(1)(1)^T(-2)(1)^T(1)^T(1)^T(1)^T\\
&~~~~~(1)(1)^T(1)(1)(1)(1)(-2)^T(1)(1)^T(1) \\
\\
\tilde{B}_{\rm sweep}& = (1)^T(2)(1)^T(-3)^T(2)(2)(1)(2)^T(2)^T \\
&~~~~~(-3)(2)^T(1)(1)(1)^T\\
\\
\tilde{C}_{\rm sweep}& = (1)^T(2)(3)^T(1)^T(-4)(3)^T(3)(-4)^T \\
&~~~~~(1)(3)(2)^T(1).
\end{split}
\end{equation}

\begin{figure}[t!]
  \includegraphics[width=8.5cm]{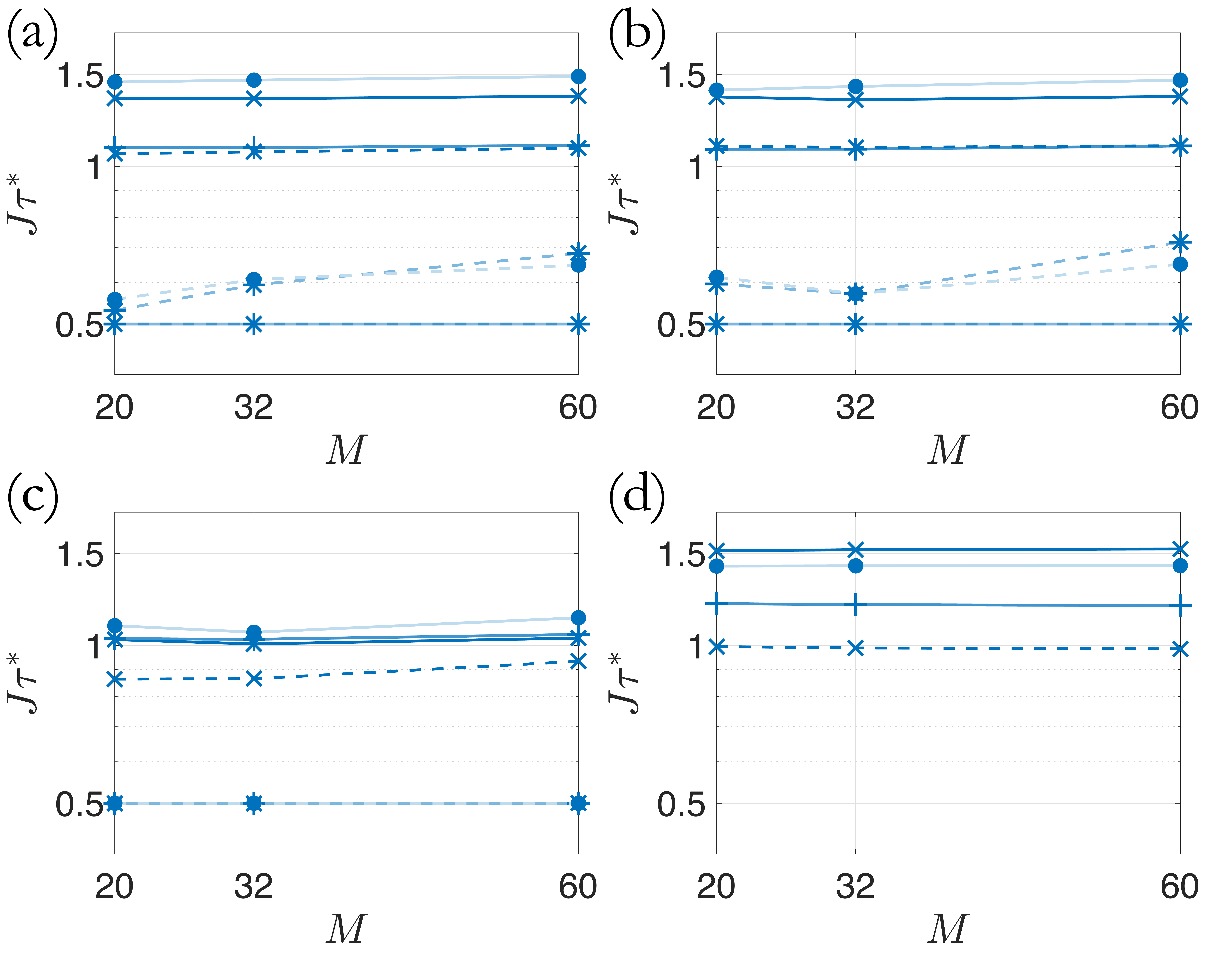}
  \centering
  \caption{Comparison of errors due to fourth order Totter decompositions to errors arising from a calibration uncertainty in the Hubbard model. Shown are the critical time-steps, $J\tau^*$, at which a digital simulation would come with Trotter errors that match the errors in an analogue simulation due to a calibration error of $0.1\%$ at $tJ = 3$. We compare a left right sweep decompositions, $\tilde{A}_{\rm sweep}$ (crosses), $\tilde{B}_{\rm sweep}$ (plus signs), $\tilde{C}_{\rm sweep}$ (stars) defined in Eq.~\ref{equ_4th_decomp_LR} to an odd-even Suzuki-Trotter split (circles) defined in Eq.~\ref{equ_4th_decomp_OE}. Additionally, we compare the cases where each two site gate contains both the tunnelling and interaction terms (solid) to the case where these pieces are also split into separate time-evolution operations (dashed). (a) Errors in $O_n = c^{\dagger}_{n,\uparrow} c_{n,\uparrow} $. (b) $O_n = c^{\dagger}_{M/2,\uparrow} c_{n,\uparrow} $. (c) $O_n = c^{\dagger}_{M/2,\uparrow} c_{M/2,\uparrow} c^{\dagger}_{n,\uparrow} c_{n,\uparrow} $. (d) Error in the state fidelity (Eq.~\ref{equ_error_fid}).}
  \label{SM_Fig11}
  \end{figure}

We also consider a fourth order Suzuki-Trotter odd-even decomposition derived in Ref.~\cite{SCHOLLWOCK201196}, which is defined by,
\begin{equation} \label{equ_4th_decomp_OE}
\tilde{U}_{\rm 4th~OE} = \tilde{U}_{\rm OE}(\tau_1)\tilde{U}_{\rm OE}(\tau_2)\tilde{U}_{\rm OE}(\tau_3)\tilde{U}_{\rm OE}(\tau_2)\tilde{U}_{\rm OE}(\tau_1),
\end{equation}
where $\tilde{U}_{\rm OE}(\tau_i)$ is given by the second order odd-even decomposition in Eq.~\ref{equ_2nd_decomp_OE}, $\tau_1 = \tau_2 = \tau/(4 - 4^{1/3})$ and $\tau_3 = \tau - 4\tau_1$.

In the same way as before we compute the critical time step, $\tau^*$, at which point the errors in an observable at $tJ=3$ become equal to the errors that arise from a calibration error in an analogue simulation. We plot the results in Fig.~\ref{SM_Fig11} as a function of system size for each of the fourth order decompositions that we have considered, where we again compare to with an without an additional splitting of the interaction and tunnelling terms. 

We find that this further splitting does not lead to large corrections and that there is only a weak dependence on system size for the time scales considered. For these fourth order decompositions, we find that for these observables that the left-right sweep decompositions are optimal which is in contrast to the second order version where we found a slight improvement if we use the odd-even decomposition. It should however be taken into account that the odd-even decomposition requires fewer steps to execute a single time step compared to any of the sweep decompositions and so when calculating the numbers of gates needed to execute a simulation with comparable error to analogue we should find the combination of step count and critical time step that results in the lowest total number of gates.

\subsection{Transverse Ising Model}

Here we present the results for a digital simulation of the transverse Ising model (see Eq.~\ref{Ising_Mod}) with algebraically decaying interactions ($\alpha=2$). In this case, it is not as straightforward to define an odd-even decomposition due to the additional presence of longer distance interactions so we only consider the case of a left-right sweep. 

In this case we can decompose our Hamiltonian into a piece that acts on two sites $H_{n,m}$, which are not necessarily on nearest neighbours. Thus we have to define our individual sweeps slightly differently, 
\begin{equation}
\begin{split}
(s) &= \prod_{n=1}^{M-1} \prod_{m=n+1}^{M}  e^{-i H_{n,m} s\tau/2} \\
&= \prod_{n=1}^{M-1} \prod_{m=n+1}^{M} \tilde{U}_{n,m}(s\tau/2).
\end{split}
\end{equation}
This means that we first apply $\tilde{U}_{1,2}$, then $\tilde{U}_{1,3}$ until $\tilde{U}_{1,M}$ before moving on to $\tilde{U}_{2,3}$, then $\tilde{U}_{2,4}$ and so on. The application of a two site time-evolution operator in the MPS algorithm on sites that were not nearest neighbour was facilitated by applying sequences of swap gates. In the resource estimation for digital quantum computation all-to-all coupling is assumed so the swap gates are not counted but would need to be added to the calculation for a architecture with less connectivity.

\begin{figure}[t]
  \includegraphics[width=8.5cm]{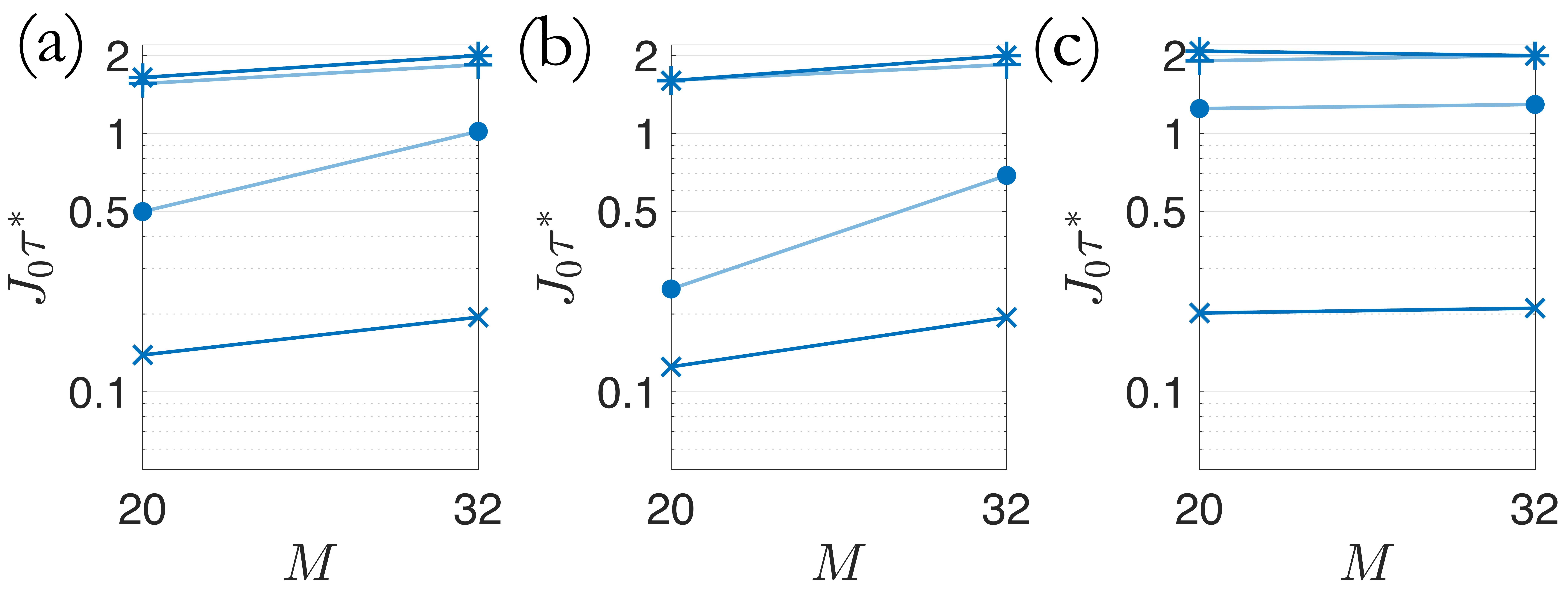}
  \centering
  \caption{Comparison of errors due to second and fourth order Totter decompositions to errors arising from a calibration uncertainty in the long-range Transverse Ising model ($\alpha=2$). Shown are the critical time-steps, $J\tau^*$, at which a digital simulation incurs Trotter errors that match the errors in an analogue simulation due to a calibration error of $0.1\%$ at $tJ_{0} = 4$. We compare the second order left-right sweep decomposition (crosses) defined in Eq.~\ref{equ_2nd_decomp_LR} to the set of fourth order left-right sweep decompositions, $\tilde{A}_{\rm sweep}$ (plus signs), $\tilde{B}_{\rm sweep}$ (stars), $\tilde{C}_{\rm sweep}$ (circles) defined in Eq.~\ref{equ_4th_decomp_LR}.(a) Errors in $O_n = s_x^n $. (b) $O_n = s_+^{M/2} s_-^n $. (c) $O_n = s_z^{M/2} s_z^n $.}
  \label{SM_Fig12}
  \end{figure}

This then allows us to define the second order decomposition using Eq.~\ref{equ_2nd_decomp_LR} and then the fourth order version using Eq.~\ref{equ_4th_decomp_LR}.

As with the Hubbard model we perform a time-evolution simulation using these decompositions for different time-steps, where we begin in an initial product state, $|\Psi(0)\rangle = | +,+,+,+\cdots\rangle$ and use the parameters $B=J_{0}$ and $\alpha=2$. We then calculate the errors which arise from this finite time-step for the different decompositions at $tJ_{0} = 4$ and extrapolate to find the ctirical time-step $\tau^*$ at which these errors are the same as those resulting from calibration error.

In Fig.~\ref{SM_Fig12} we plot the results from the second and fourth order decompositions, where we find similar behaviour to the Hubbard model where the values do not strongly depend on system size, but are more sensitive to the form of the decomposition used. 

\section{Digital resource estimation}

\label{app:gatecount}

In this appendix we will explain in more detail the methods used to calculate and minimise the resources required to execute a digital quantum simulation of comparable accuracy to that of an analogue quantum simulator. As we assume that the limiting factor in the accuracy of the digital simulation is the error due to Trotterisation, we need take as inputs only the number of sweeps required per time step and the total number of time steps. These depend on the choice of Trotter decomposition, the duration of a time step, $\tau$, and the total duration of the simulated evolution, $t$. By finding the number of gates and their depths per sweep we are then able to report the total number of gates and gate depths for the full evolution. The true optimal decomposition will depend also on the gate errors but by choosing to assume perfect gates we are nevertheless able to illustrate the methods one would use to minimise gate counts, report on the gate counts for a fault-tolerant system, and give optimistic estimates for a NISQ system.

\subsection{Hubbard Model}

For the Hubbard model we use 2 qubits to represent each site, where each qubit represents one fermionic species. The Jordan-Wigner mapping \cite{jordan1928, whitfield10} ensures that the fermionic anticommutation relation is obeyed which yields the following mappings:
\begin{equation}\label{eq:chemical_potential}
n_{k} = \frac{1}{2}(I - \sigma_{z}^k) \rightarrow -\frac{1}{2} \sigma_{z}^k,
\end{equation}
\begin{equation}\label{eq:repulsion}
(n_{k} - \frac{1}{2})(n_{l} - \frac{1}{2}) =\frac{1}{4}\sigma_z^k\sigma_z^l,
\end{equation}
\begin{equation}\label{eq:hopping}
c^{\dagger}_{k}c_{l} + h.c. = \frac{1}{2}(\sigma_x^k \sigma_x^l + \sigma_y^k \sigma_y^l)Z_{JW},
\end{equation}
where $Z_{JW}$ represents the tensor product of $\sigma_{z}$ on the sites between $k$ and $l$, the so called Jordan Wigner strings,
\begin{equation}
Z_{JW} = \sigma_z^{k+1}\sigma_z^{k+2}...\sigma_z^{l-1}
\end{equation}
and where $k$ and $l$ index both site and species of the Hubbard model and hence map to a single qubit. The inclusion of a chemical potential term in the Hamiltonian would incur no extra cost due to the necessity of single qubit operations to correct the execution of the on-site interaction term.

\begin{figure}[t]
  \includegraphics[width=8.5cm]{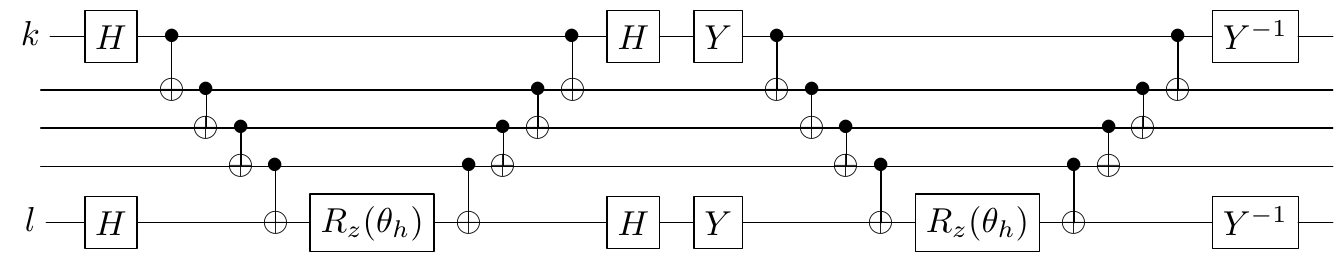}
  \centering
  \caption{Circuit diagram showing implementation of the tunneling term of the Hubbard model between qubits $k$ and $l$. Note that here the $Y$ and $Y^{-1}$ gates describe the operations that execute and reverse a basis change from $z$ to $y$ respectively, as $H$ converts $z$ to $x$ and is its own inverse. $\theta_h = -J$.}
  \label{SM5_Fig1}
\end{figure}

If we assume a native gate set of Clifford gates, CNOTs and an arbitrary rotation of angle $\theta$ clockwise around the $z$ axis defined by $R_z(\theta) = e^{-i\theta\sigma_z/2}$ then executing the the terms of Eq.~\ref{eq:chemical_potential} and \ref{eq:repulsion} involve single qubit rotations and two CNOT gates per term but is otherwise trivial. The tunnelling term is more complex due to the presence of the Jordan Wigner string. The naive circuit used to implement a single tunneling term is shown in Fig.~\ref{SM5_Fig1}. Three main observations can be used to reduce the number of CNOT gates needed to execute multiple such terms and parallelise the rotation gates as much as possible \cite{hastings15}.

\begin{figure}[t]
   \includegraphics[width=8.5cm]{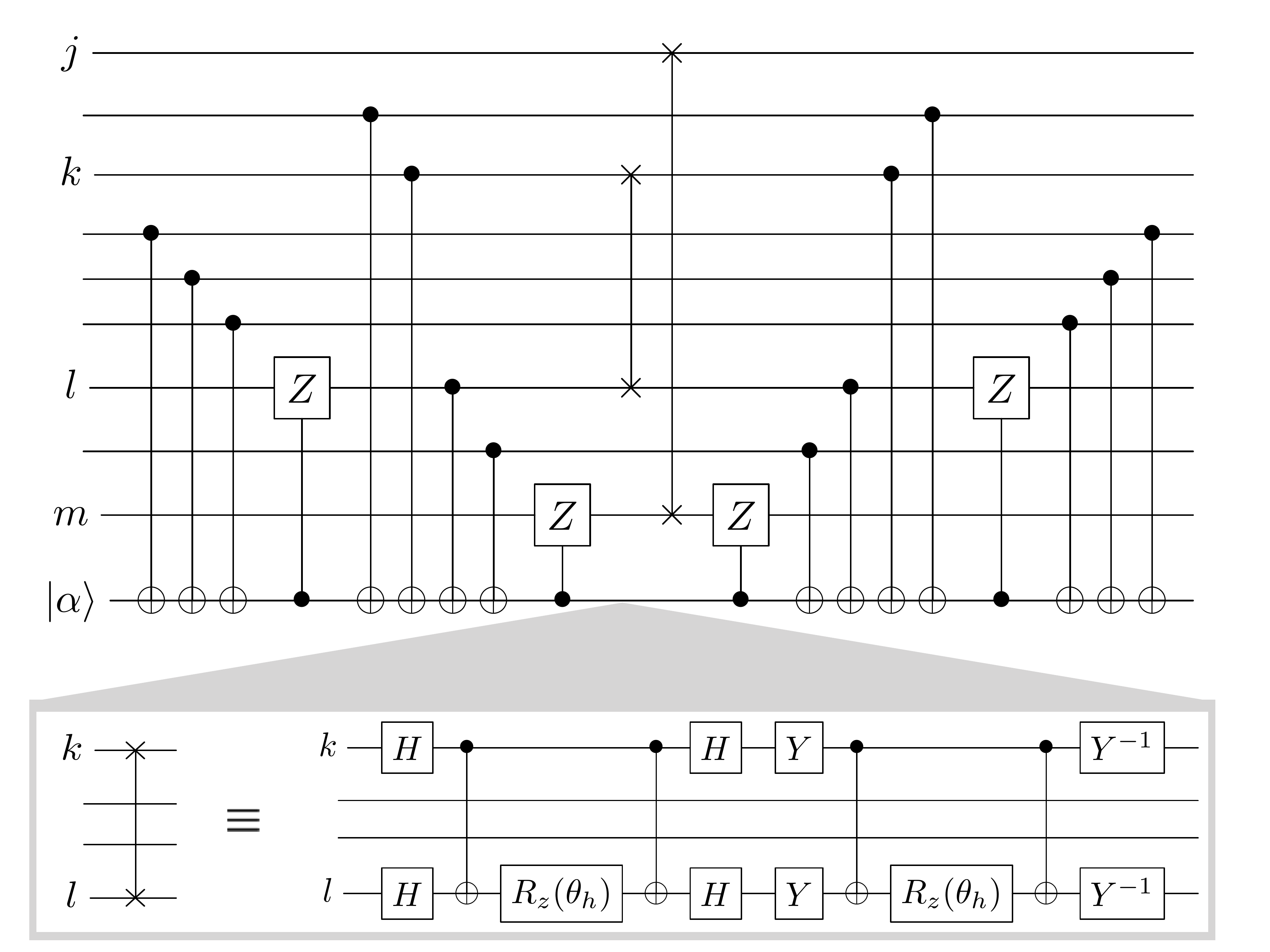}
  \centering
  \caption{Circuit diagram showing the optimised execution of two tunnelling events between qubits $k$ and $l$ and $j$ and $m$, where the parity calculation between $k$ and $m$ is computed onto an ancilla qubit $\alpha$ and reused in the $j$ to $m$ tunelling computation. The shorthand used to describe the rotation parts of the circuit is shown in the lower panel. Notably this rotation part is the complete optimised hopping term between $k$ and $l$in the case of no intervening qubits (i.e. $l = k + 1$)}
  \label{SM5_Fig2}
\end{figure}

Firstly, since the Jordan Wigner string simply counts the number of excitations on the intervening qubits, the ordering of the operations used to calculate the string can be changed such that the calculated parity used to execute the rotation around the $x$ axis can be reused to execute the rotation around the $y$ axis. This could also be seen by permuting the basis changing $H$ and $Y$ gates through the CNOT gates and cancelling the CNOT gates that are now adjacent.

Secondly, again exploiting our understanding of the purpose of the Jordan Wigner string, it is preferable when doing multiple parity calculations to store the calculated parity on an ancilla qubit for re-use when calculating longer strings. A calculation of the parity between $k$ and $l$ can be re-used when calculating the string between $j$ and $m$ given $j > k > l > m$.

The third observation is that tunneling between two sites does not affect the parity between these two sites and so can be commuted through an encompassing string, thus allowing the rotation part of the two terms to be executed in parallel. Fig.~\ref{SM5_Fig2} shows the circuit which executes two such tunneling events between $j$ and $m$ and between $k$ and $l$, having implemented all optimisations. 

\begin{figure}[t]
   \includegraphics[width=5cm]{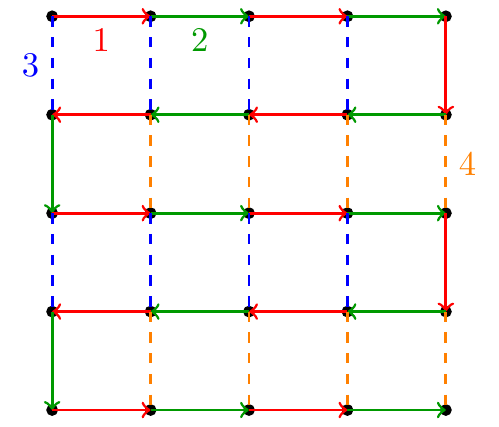}
  \centering
  \caption{Lattice qubit ordering and hopping term groups for a qubits representing one spin species on a 2D lattice with $M=25$ sites. Black circles represent the qubits, hopping terms which can be executed in parallel are grouped by colour, and solid (dashed) lines represent hopping terms between neighbouring (distant) qubits by index.}
  \label{SM5_Fig3}
\end{figure}
 
As well as minimising the number of CNOT gates and parallelising the rotations for nested tunneling events it is beneficial to execute as many operations in parallel as possible. This is achieved by ordering the qubits first by the species of fermion which they represent and within which tunnelling events may take place. Each group is arranged in a 2D lattice matching layout of the corresponding sites as shown in Fig.~\ref{SM5_Fig3}. The qubits within each lattice are numbered using a snake-like configuration such that hopping events between qubits neighbouring in the horizontal direction need no Jordan Wigner strings. All of these events can be executed in just two steps, where within each step all elements commute as they do not share qubits. Tunnelling events in the vertical direction are achieved by executing two further steps, first starting with qubits on even rows and then on odd ones. Jordan Wigner strings spanning the row in order to execute these vertical interactions will not all commute with one another but will be nested so one can re-use the parity calculations as described above to minimise the number of gates.

Thus we are able to count the necessary gates to implement a single sweep of evolution under the Hubbard model, shown in Tab.~\ref{tab:hubbardcounts} and where, for this arrangement of qubits and choice of optimisation, one ancilla per row will be needed, requiring $\sqrt{M}$ extra qubits.

\begin{table}
\centering
\begin{tabular}{ l | c | r }
  Gate & Gate Count & Depth\\ \hline 
  Cliffords (H, Y) & $32(M - \sqrt{M})$ & 16\\
  $R_z(\theta)$ & $11M - 8\sqrt{M}$ & 10\\
  CNOT &  $2(15M-20\sqrt{M}+6)$ & $6(2\sqrt{M} + 1)$ \\
\end{tabular}
\caption[Total resources required to implement single time-step-sweep of the Fermi-Hubbard model on a lattice]{Total resources required to implement single time step sweep for an $L\times L$ lattice with $M$ sites of the Hubbard model.}\label{tab:hubbardcounts}
\end{table}

Combining these numbers with the number of sweeps per time step and the total number of time steps results in the gate counts found in the main text. In the likely case that it is not possible to execute such large numbers of gates on a NISQ device we must use a fault-tolerant system. The main effect in terms of resource estimation is that all rotation gates must be constructed using many $T$ gates following \cite{bocharov15}, but where savings can be made by executing the rotations in parallel as described in \cite{gidney18}.

\subsection{Transverse Ising Model}

The Ising model is comparatively simple to execute, requiring only as many qubits as sites and with the transverse term mapping trivially to a single qubit rotation around the $x$ axis (and any optional longitudinal term mapping to a $z$ axis rotation). The more complex coupling term will require a fixed number of gates but the depth of these can be optimised in order to execute as many gates as possible simultaneously. We have here assumed all-to-all coupling, motivated by the increased complexity of this interaction landscape. The resulting gate counts and depths per sweep for the full Hamiltonian (without longitudinal term) are shown in Tab.~\ref{tab:isingcounts}

\begin{table}[t]
\centering
\begin{tabular}{ l | c | r }
  Gate & Gate Count & Depth\\ \hline 
  Cliffords (H) & 2M & 2 \\
  $R_z(\theta)$ & $M + M(M//2)$ & $M+2$\\
  CNOT &  $2M(M//2)$ & 2M \\
\end{tabular}
\caption{Total resources required to implement single time step sweep of the long-range transverse field Ising model with M sites.}\label{tab:isingcounts}
\end{table}

In the main text we again evaluate this for a given Trotter decomposition, time step size and total evolution duration to give the total resource requirements and then find the fault-tolerant estimates based on the same methods and exploiting the parallel execution of rotation gates to minimize the number of $T$ gates needed. 

\bibliographystyle{apsrev}
\bibliography{QA_QS}

\end{document}